\documentclass[]{aastex631}

\usepackage{graphicx}
\usepackage{color}
\usepackage{epsfig}
\usepackage{subfigure}
\usepackage{amssymb}
\usepackage{amsmath}
\usepackage{natbib}
\usepackage{tabularx}
\usepackage{rotating}
\usepackage{multirow}
\usepackage{ulem}
\usepackage{longtable}
\usepackage{threeparttable}
\shorttitle{Classification of Periodic Variable Stars from TESS}
\shortauthors{X. Gao, X. Chen, S. Wang, and J. Liu}
\graphicspath{ {Figures/} }
\begin{document}

\title{Classification of Periodic Variable Stars from TESS}

\author{Xinyi Gao}
\affiliation{CAS Key Laboratory of Optical Astronomy, National Astronomical Observatories, Chinese Academy of Sciences, Beijing 100101, China}
\affiliation{School of Astronomy and Space Science, University of the Chinese Academy of Sciences, Beijing, 100049, China}

\author{Xiaodian Chen}
\email{chenxiaodian@nao.cas.cn}
\affiliation{CAS Key Laboratory of Optical Astronomy, National Astronomical Observatories, Chinese Academy of Sciences, Beijing 100101, China}
\affiliation{School of Astronomy and Space Science, University of the Chinese Academy of Sciences, Beijing, 100049, China}

\author{Shu Wang}
\affiliation{CAS Key Laboratory of Optical Astronomy, National Astronomical Observatories, Chinese Academy of Sciences, Beijing 100101, China}

\author{Jifeng Liu}
\affiliation{CAS Key Laboratory of Optical Astronomy, National Astronomical Observatories, Chinese Academy of Sciences, Beijing 100101, China}
\affiliation{School of Astronomy and Space Science, University of the Chinese Academy of Sciences, Beijing, 100049, China}

\begin{abstract}
The number of known periodic variable stars has increased rapidly in recent years. As an all-sky transit survey, the Transiting Exoplanet Survey Satellite (TESS) plays an important role in detecting low-amplitude variable stars. Using 2-minute cadence data from the first 67 sectors of TESS, we find 72,505 periodic variable stars. We used 19 parameters including period, physical parameters, and light curve (LC) parameters to classify periodic variable stars into 12 sub-types using random forest method. Pulsating variable stars and eclipsing binaries are distinguished mainly by period, LC parameters and physical parameters. GCAS, ROT, UV, YSO are distinguished mainly by period and physical parameters. Compared to previously published catalogs, 63,106 periodic variable stars (87.0$\%$) are newly classified, including 13 Cepheids, 27 RR Lyrae stars, $\sim$4,600 $\delta$ Scuti variable stars, $\sim$1,600 eclipsing
binaries, $\sim$34,000 rotational variable stars, and about 23,000 other types of variable stars. The purity of eclipsing binaries and pulsation variable stars ranges from 94.2$\%$ to 99.4$\%$ when compared to variable star catalogs of Gaia DR3 and ZTF DR2. The purity of ROT is relatively low at 83.3$\%$. The increasing number of variables stars is helpful to investigate the structure of the Milky Way, stellar physics, and chromospheric activity.
\end{abstract}
\section{Introduction} \label{sec:intro}

Periodic variable stars have been studied for centuries. Periodic variable stars consist mainly of pulsation stars, eclipsing binary systems and rotational stars that exhibit regular or semi-regular luminosity variations. The periods of variable stars are usually associated with their physical properties, such as mass, luminosity, radius, and ages \citep{2018ApJS..237...28C}. Variable stars have long been used to provide important insights into stellar structure and evolution \citep{2015ApJ...808...50M, 2019AA...623A.110G, 2024ApJS..274...30Z}. Variable stars are not only important objects for constraining stellar evolution, but also important objects for studying the Milky Way structure \citep{2019NatAs...3..320C, 2019Sci...365..478S}. The period--luminosity relations (PLRs) of classical Cepheids and RR Lyrae, for example, were used in the cosmological distance scale \citep{2001ApJ...553...47F, 2023NatAs...7.1081C, 2022ApJ...934L...7R}.

The study of variable stars dates back hundreds of years, but in the 1990s the development of CCDs led to significant advances in variable star search. A large number of large-scale surveys, including ground-based surveys and space-based surveys, have also contributed to the development of variable star discovery. The All Sky Automated Survey (ASAS) made the first variable star search covering almost the entire sky \citep{1997AcA....47..467P}. ASAS found 3,126 variable stars in the V band throughout the southern hemisphere and 11,509 variable stars brighter than $V = 15$ mag throughout the northern hemisphere \citep{2002AcA....52..397P, 2005AcA....55..275P}. The Optical Gravitational Lensing Experiment (OGLE) has been used to study the variable stars for more than two decades \citep{2008AcA....58...69U}. It discovered 900,000 variable stars and led to significant advances in variable star research \citep{1992AcA....42..253U, 2015AcA....65....1U}. \cite{2022MNRAS.513..420B} extracted variability periods for 16,880 stars in the Next Generation Transit Survey (NGTS). The Gaia second Data Release (DR2) mapped all-sky RR Lyrae stars and Cepheids and discovered 50,570 new variable stars \citep{2019A&A...622A..60C}. \cite{2020ApJS..249...18C} searched and classified 781,602 periodic variable stars from Zwicky Transient Facility (ZTF) DR2. The All-Sky Automated Survey for Supernovae (ASAS-SN) obtained 687,695 variable stars throughout the sky, including 238,752 new discoveries \citep{2014ApJ...788...48S,2018MNRAS.477.3145J,2019MNRAS.486.1907J,2019MNRAS.485..961J,2020MNRAS.491...13J,2021MNRAS.503..200J,2023MNRAS.519.5271C}. In Gaia DR3, 12.4 million variable stars in the Milky Way and nearby galaxies were classified by supervised machine learning into 22 variability types \citep{2023A&A...674A..14R}.

As space telescopes, Kepler and Transiting Exoplanet Survey Satellite (TESS) play an important role in variable source searches for high photometry precision. The Kepler spacecraft, launched into an Earth-trailing orbit, focused almost continuously on a single region of the sky for four years, monitoring the brightness fluctuations of approximately 200,000 stars \citep{2010Sci...327..977B,2018ApJS..235...38T}. These data have facilitated the study of stellar variability in an unprecedented way. \cite{2009AcA....59...33P} found 947 variable stars in the Kepler field of view. \cite{2011AJ....141...83P} found 1,879 eclipsing binaries in the first Kepler data release, and \cite{2011AJ....142..160S} found 2,165 eclipsing binaries in the second Kepler data release. \cite{2011AJ....141...20B} provided an overview of stellar variability in the first quarter data from the Kepler mission. Following the failure of the Kepler mission reaction wheels, the K2 mission was initiated and observed both the northern and southern sky over two years, covering ten times more sky area than the original Kepler mission \citep{2014PASP..126..398H}. However, K2 experiences higher instrument noise due to increased instability of the spacecraft. TESS aims to discover planets by observing transits across bright and nearby stars, and an all-sky survey approach was adopted for TESS \citep{2015JATIS...1a4003R,2018AJ....156..102S}. \cite{2018AJ....156..102S} and \cite{2019AJ....158..138S} described the TESS Input Catalogs (TIC). \cite{2019MNRAS.485..961J} identified $\sim$11,700 variables, including $\sim$7,000 new discoveries.  \cite{2020AJ....159...60G} studied stellar flares for the 24,809 stars observed with a 2 minute cadence during the first two months of the TESS mission and identified 1,228 flaring stars. \cite{2020MNRAS.493.1388Y} measured the pulsation amplitude and period for 3,213 long period variable stars. \cite{2020MNRAS.499.5508S} found 1,807 variable stars in full frame images. \cite{2021ApJS..254...39G} presented 2,241 exoplanet candidates identified with data from TESS during its 2 yr Prime Mission. \cite{2021MNRAS.503.3828B} found 506 variable objects. In the northern TESS continuous viewing zone, 3,025 A-F variable stars brighter than 11 mag were found, of which 1,813 objects were classified \citep{2022A&A...666A.142S}. \cite{2023MNRAS.519.2486S} identified 1,403 variable stars and \cite{2023ApJS..268....4F} found over 46,000 periodic objects with high confidence. Specific types of variable stars have also been discovered in TESS observations, e.g. more than 20,000 binaries and 307 Ia supernovae \citep{
2023MNRAS.525.4596D, 2023ApJ...956..108F, 2023MNRAS.522...29G, 
2022RNAAS...6...96H, 2022ApJS..258...16P}. \cite{2023ApJS..268....4F} presented a stellar variability catalog using the 2 minute cadence photometry obtained during the TESS 2yr Prime Mission. There is still a dearth of work on the careful classification of large samples of variable stars in the TESS. 

In this work, we search for periodic variable stars in 67 sectors of the TESS and use the machine learning method and classify 70,100 periodic variable stars into 12 sub-types.
%19 physical parameters that are important for their corresponding variables classification were analysed. The rate of new found variables in our sample is up to 79.5\%, beneficial to further studies of variable in the future. 
The data are described in Section \ref{sec:TESS}. In Section \ref{sec:Search for Periodic Variable Stars}, we show how to identify periodic variable stars. We classify them using a machine learning method and show how the parameters work in classification in Section \ref{sec:Classification for Periodic Variable stars}. Section \ref{sec:The Periodic Variable stars Catalog} shows the variable star catalog, the newly classified variable stars, and a consistency comparison with the two previous catalogs. We describe and compare the properties of each subtype of variable star in detail in Section \ref{sec:Discussion of Periodic Variable stars}. Section \ref{sec:Conclusions of Periodic Variable Stars} summarizes our conclusions.

\section{TESS data} \label{sec:TESS}
\begin{figure*}
\centering
\begin{minipage}{185mm}
  \includegraphics[width=185mm]{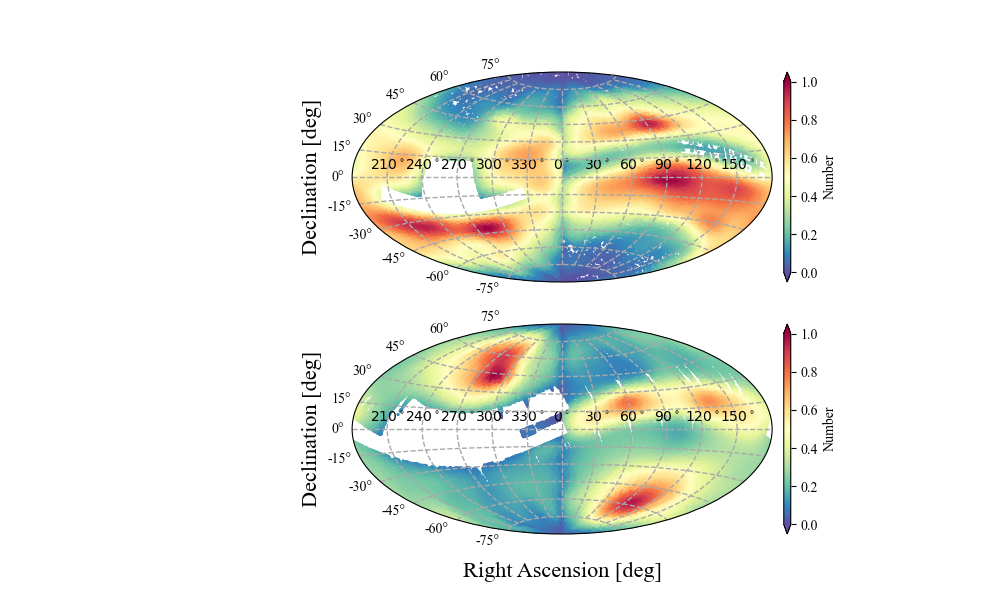}
\caption{Distribution in equatorial coordinates of stars observed at 2 minute cadence by TESS in 67 sectors. Upper: stars that were observed in only one sector of TESS observations (27.4 days). Lower: stars that were observed in multiple sectors of TESS observations
($>54.8$ days). The number in the color bars is the normalized number.}
\label{Figure1}
  \end{minipage}
\end{figure*}

As an astrophysics exploration mission, TESS is placed in a highly elliptical 13.7-day orbit around the earth and uses four wide-field optical CCD cameras to observe a $24^\circ\times96^\circ$ region of the sky. Compared to ground-based telescopes, TESS can obtain data for low amplitudes objects. The time-series photometry is obtained in 27.4 day segments, known as sectors. It recorded brightness measurements of selected target stars every 2 minutes and full-frame images every 30 minutes. The goal of TESS is to discover and characterize planets around nearby and bright stars \citep{2014SPIE.9143E..20R, 2015JATIS...1a4003R}. Only data from the first 67 sectors of TESS 2-minute cadence photometry are used in this work. They belong to Prime Mission, Extended Mission, and Extended Mission 2 up to 2023. The upper panel of Figure \ref{Figure1} shows the
distribution of stars observed in a single sector, while the lower panel of Figure \ref{Figure1} shows the distribution of stars observed in multiple sectors. Some of the data presented in this work were obtained from the Mikulski Archive for Space Telescopes (MAST) at the Space Telescope Science Institute. The specific observations analyzed can be accessed through \dataset[doi:10.17909/t9-nmc8-f686]{https://archive.stsci.edu/doi/resolve/resolve.html?doi=10.17909/t9-nmc8-f686}.

\section{Search for Periodic Variable Stars} \label{sec:Search for Periodic Variable Stars}
To search for periodic variable stars quickly, we need to do some pre-processing to make the search process faster. We divided the objects in 67 sectors into two main groups: 249,482 objects observed in only one sector and 228,477 objects observed in multiple sectors. For objects observed in multiple sectors, we normalized the presearch data conditioning SAP (PDCSAP) flux by dividing the PDCSAP fluxes of the same object in different sectors by their average \citep{2016SPIE.9913E..3EJ}.

We then used \textit{astroperies.timeseries.LombScargle} function in python to determine the period. The inverse of the frequency corresponding to the highest power peak in the periodogram is the optimal period. A sector with a time span of 27.4 d may not be able to contain two full cycles for long-period objects, so the uncertainty of the period is high. Therefore, the periods of objects observed in only one sector were limited from 0.01 to 13 days and the periods of objects observed in more than one sector were limited from 0.1/$\sqrt{N}$ (N is number of sectors) to 13$\times$N days. We set 5 grid points in frequency space across each significant periodogram peak for objects observed by only one sector and 40 grid points for objects observed by multiple sectors. We also calculated the false alarm probability (FAP), which represents the confidence of the period in the LC. It can be used to select periodic variable stars. 

We estimated from our simulations how many false periodic variable stars would be selected by the FAP criterion. The PDCSAP LCs remove the common instrumental systematics, correct for the amount of flux captured by the photometric aperture and crowding from known nearby stars\footnote{\url{https://outerspace.stsci.edu/display/TESS/2.0+-+Data+Product+Overview}}. Thus, we only need to consider the effect of noise on the PDCSAP flux. We simulated the effect of noise on the FAP for single and multiple sectors separately to determine how much of the noise data was misidentified as periodic variable stars. For objects observed in a single sector, we randomly selected 1000 targets and generated 1000 sets of noise data per target matching the length of the PDCSAP flux, for a total of 1,000,000 sets of noise data to ensure statistical significance. The mean value of the simulated LC ($\mu$) was consistent with the mean value of the PDCSAP flux, while the standard deviation of the added noise ($\sigma$) was consistent with the mean value of the PDCSAP flux error. The flux and flux error for each target were estimated from their original LCs. We then determined the period and FAP for these simulated LCs. The analysis shows that only 62 sets of noisy data have a FAP lower than 0.001. This means that for every million objects observed in a single sector, about 62 sets could be misclassified as periodic variable star candidates due to the noise. For objects observed in multiple sectors, we selected 500 objects and generated 1000 sets per target, for a total of 500,000 noise data sets. Out of the 500,000 noisy data sets, about 117 sets could be incorrectly identified as periodic variable star candidates. 

In this work, we use a FAP of less than 0.001 to select periodic variable star candidates. We can infer that about 10 of the 166,163 objects observed in a single sector and 45 of the 191,910 objects observed in multiple sectors may have been incorrectly identified as periodic variable star candidates. Whether these objects are periodic or not needs to be determined by more TESS observations.

We converted PDCSAP fluxes to magnitudes for all objects. For objects observed by multiple sectors, we subtracted the mean magnitude of the different sectors from the magnitude, i.e., we zeroed out the mean magnitude of each observed sector. We excluded outliers that deviated from the mean by more than $\textgreater$5$\sigma$ and fitted the LC with an eighth-order Fourier function $f = a_0 + \sum_{i=1}^{8} a_i \cos(2\pi it/P + \phi_i)$. $a_i$ and $\phi_i$ refer to the amplitude and phase of each order Fourier function, respectively. The higher order function can better fit more complex LCs. We found that using the eighth-order Fourier function gives a better fit than using the fourth-order Fourier function in fitting LCs of RRab, RRc, T2CEP, DCEP, EA, and EB. Better fitting parameters are important for variable star classification using machine learning method.

\begin{deluxetable*}{cccc} \tablecaption{ Purity test of periodic variable stars at different $R^2$ thresholds.\label{Table1}} \tablehead{\colhead{$R^2$ thresholds.}& \colhead{Number of periodic variable stars}& \colhead{Total Number}&\colhead{Purity}} 
\startdata 
0.05&106,530&132,896&80.1\%\\
0.10&82,054&92,164&89.0\%\\
0.13&81,130&88,184&92.0\%\\
0.40&32,656&32,780&99.6\%\\
\enddata 
\end{deluxetable*}

Before proceeding with the classification of variable stars, we need to exclude poorly fitted variable stars from the periodic variable star candidates, most of which have low signal-to-noise ratios. $R^2$ represents how well the Fourier model fits the LCs. A higher $R^2$ indicates a better fit. We set a threshold for $R^2$ to distinguish between variable stars and non-variable stars. Using a smaller $R^2$ threshold will result in more non-periodic variable stars being mixed in with the periodic variable star candidates, and conversely, using a larger $R^2$ threshold will result in a smaller sample of periodic variable star candidates. We investigate the effect of different $R^2$ on the purity of final periodic variable stars through a random forest model test.

We added 500 non-periodic variable stars with $R^{2} \approx 0$ (labeled as `non-variable') in the training set (about 25\% of the training set, see Section \ref{sec:Search for Periodic Variable stars}), with others labeled as `variable' to train a classifier. The non-periodic variable star sample includes 300 objects observed by a single sector and 200 objects observed by multiple sectors. Using the training set, we trained a classifier to separate non-periodic variable stars and periodic variable stars. We obtained objects with a classification probability $> 0.9$ and added them back into the training set for ten iterations to refine the classifier. Finally, we applied the updated classifier to objects selected by different $R^2$ thresholds, and the results are in Table \ref{Table1}. Overall, $R^{2}=0.13$ ensures both the completeness and the purity ($>90\%$) of the periodic variable star candidates. 

We first selected variable stars with $R^2>0.13$, leaving 46,303 objects observed in only one sector and 41,881 objects observed in multiple sectors. Based on examining the LCs of objects observed in a single sector, we found that we also need to exclude objects with periods longer than 10 days unless $R^2>0.63$. Since we limited the maximum determined period for single sector observations to 13 days, when the determined period approaches this limit, boundary effects may arise, leading to inaccurate period estimates. The reason for this is that for objects with a true period greater than 13 days, a single sector observation of TESS may not capture two full periods, and thus may yield an underestimation of the period. This selection leaves 35,721 periodic variable star candidates observed in a single sector. The total number of objects is 77,602.

Next, we set the step size as $0.0001\times P$ in the range of $0.99\times P$ to $1.01\times P$ and refitted LCs with the Fourier function to obtain the optimal period corresponding to the highest $R^2$. We visually eliminated 2,449 variable stars candidates whose phase-folded LCs varied significantly from cycle to cycle, with no variation or spike-like fluctuations in some TESS cycles. Most of these excluded candidates are ROT and they have lower period and classification accuracies due to the insufficient time span of TESS. Finally, after cross-matching with Gaia DR3 and excluding objects for which absolute Wesenheit magnitudes could not be estimated, we obtained 72,505 periodic variable stars.

Overall in this section, we select periodic variable stars in four steps. First we used the Lomb–Scargle periodogram to obtain periods and FAP. 166,163 objects observed in only one sector and 191,910 objects observed in multiple sectors were selected by the FAP $\textless$ 0.001 criterion. We then fitted LCs with an eighth-order Fourier function. We excluded objects with $R^{2}$ larger than 0.13, yielding 46,303 objects observed in only one sector and 41,881 objects observed in multiple sectors. For objects observed in only one sector, we further excluded objects with periods larger than 10 days and $R^{2}$ less than 0.63, leaving 35,721 objects. Next, we checked LCs of 77,602 objects, leaving 75,153 objects. Finally, the objects were cross-matched with Gaia DR3 to extract the physical parameters, yielding a final sample of 72,505 periodic variable stars.

\begin{table*}[ht!]
% \vspace{-0.0in}
\begin{center}
\caption{\label{Table2}Variability Features.}
% \vspace{0.15in}
% \begin{tabular}{l p{8cm} c ccl}
\begin{tabular}{lccccl}
% \begin{tabular}{lccp{3cm}}
\hline
\hline
Feature&Description&Estimated by LC&Reference\\
\hline
Period  & Photometric period determined by the Lomb-Scargle &Y& \\
  & perioddogram && \\
Parallax& Photometric parallax provided by $\textit{Gaia} $DR3&&\cite{2019AA...623A.110G} \\
Error& Parallax uncertainty provided by $\textit{Gaia}$ DR3&&\cite{2019AA...623A.110G}\\
$(\textit{B}_P-\textit{R}_P)_0$  & $(\textit{G}_{BP}-\textit{G}_{RP})_0$, dereddened color determined based on &&\\
 &  $\textit{Gaia}$ DR3 &&\\
$\textit{M}_{W_G}$  & $G$-band (330 nm to 1050 nm) absolute Wesenheit  && \\
& magnitudes determined based on $\textit{Gaia}$ DR3&\\
Amp.  & Normalized peak-to-peak amplitude determined from   &Y& \\
&the eighth-order Fourier fitting of the LCs&\\
$\gamma_{2}$  & Kurtosis =$\frac{\frac{1}{n}\sum_{i=1}^{n}(x_i-\bar{x})^4}{(\frac{1}{n}\sum_{i=1}^{n}(x_i-\bar{x})^2)^2}-3$  &Y&\cite{2014566A..43K}\\
$\gamma_{1}$  & Skewness = $\frac{\frac{1}{n}\sum_{i=1}^{n}(x_i-\bar{x})^3}{(\frac{1}{n}\sum_{i=1}^{n}(x_i-\bar{x})^2)^3}/2$ & Y&\cite{2014566A..43K}\\
$Q_{31}$  & The normalized difference betweeen 3rd quartile (75\%)  &Y&\cite{2014566A..43K} \\
&and 1st quartile (25\%) in LCs&&\\
W  & Shapiro-Wilk normality test statistics & Y&\cite{2016587A..18K}\\
K  & Stetson K index, calculated using LCs &Y&\cite{1996PASP..108..851S}\\
Std  & The normalized weighted standard deviation of the LCs&Y&\\
$R_{21}$ &$a_2/a_1$, 2nd to 1st amplitude ratio obtained from &Y&\\
& the eighth-order Fourier fitting&&\\
$R_{31}$ &$a_3/a_1$, 3rd to 1st amplitude ratio obtained from &Y&\\
& the eighth-order Fourier fitting&&\\
$\phi_{21}$ &$\phi_2-2\phi_1$, the phase difference between 2nd to 1st phase &Y&\\
& obtained from the eighth-order Fourier fitting&&\\
$\phi_{31}$ &$\phi_3-3\phi_1$, the phase difference between 3rd to 1st phase &Y&\\
& obtained from the eighth-order Fourier fitting&&\\
$W_{1}-W_{3}$ & $W_{1}-W_{3}$ color determined based on $\textit{WISE}$&&\\
$W_{1}-W_{4}$ & $W_{1}-W_{4}$ color determined based on $\textit{WISE}$&&\\
$\textit{M}_{W_1}$  & The absolute $W_{1}$-band Wesenheit magnitude && \\
\hline
\end{tabular}
% \end{tabular}
\end{center}
\end{table*}

\begin{deluxetable*}{ccccc}
\tablecaption{{Acronyms for the periodic variable stars and their numbers.\label{Table3}}}
\tablehead{\colhead{Variable type}& \colhead{Acronym}& \colhead{Number in the pre-classified set}&\colhead{Predicted quantity}&\colhead{total}}
\startdata
classical Cepheids and Type-II Cepheids,&Cepheids&73&111&184\\
% Fundamental-mode classical Cepheids&DCEP&63&101&164\\
% First-overtone Cepheids&DCEPS&5&8&13\\
$\delta$ Scuti variable stars&DSCT&59&4828&4887\\
detached eclipsing binaries&EA&542&3396&3938\\
semi-detached eclipsing binaries&EB&111&293&404\\
contact eclipsing binaries&EW&282&1019&1301\\
$\gamma$ Cassiopeiae variable stars&GCAS&88&23553&23641\\
high amplitude $\delta$ Scuti variable stars&HADS&57&127&184\\
Rotational variable stars&ROT&709&36031&36740\\
fundamental-mode RR Lyrae variable stars&RRab&291&42&333\\
first-overtone and double-mode RR Lyrae variable stars&RRcd&108&31&139\\
% First-overtone RR Lyrae variable stars&RRc&105&28&133\\
% Double-overtone RR Lyrae variable stars&RRd&3&3&6\\
% Type-II Cepheids&T2CEP&5&2&7\\
% U Geminorum-type variable stars (dwarf novae)&UG&8&84&92\\
eruptive variable stars of the UV Ceti type&UV&45&432&477\\
Young stellar objects&YSO&40&237&277\\
sum&&2405&70100&72505\\
\enddata
\end{deluxetable*}

\begin{figure*}
\centering
\begin{minipage}{185mm}
  \includegraphics[width=185mm]{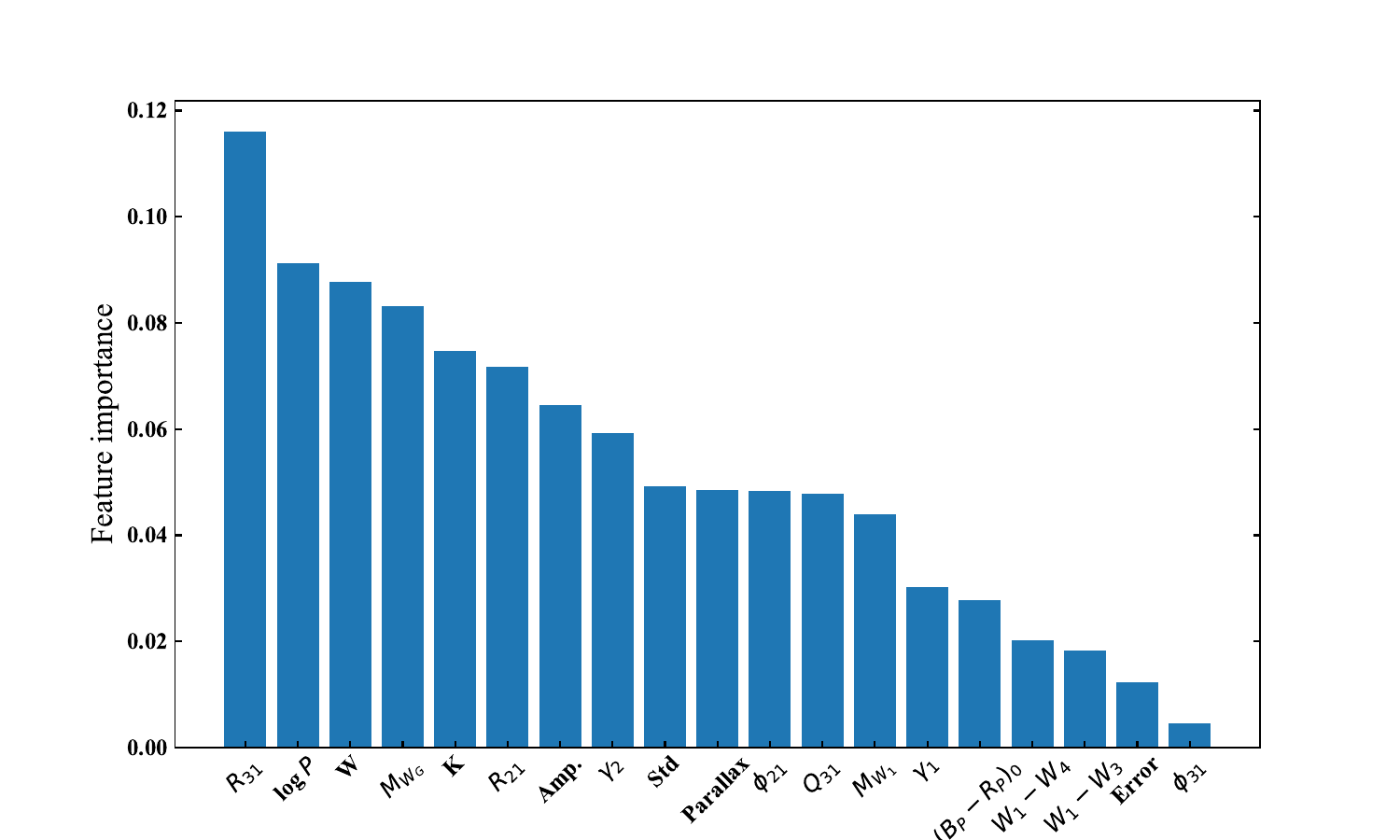}
\caption{{Feature importance as estimated using the random forest algorithm. Obviously, $R_{31}$ is the most powerful feature for separating periodic variable stars. The importance of period is also very high, only second to $R_{31}$.}}
\label{Figure2}
  \end{minipage}
\end{figure*}

\begin{deluxetable*}{ccccc}
\tablecaption{{Classification quality of the trained model.\label{Table4}}}
\tablehead{\colhead{Type}& \colhead{Precision}& \colhead{Recall}&\colhead{$F_1$-score}&\colhead{Number}}
\startdata
% DCEP&0.93&1.00&0.96&63\\
% DCEPS&1.00&0.20&0.33&5\\
Cepheids&1.00&0.96&0.98&24\\
DSCT&0.95&0.95&0.95&19\\
EA&0.99&0.98&0.99&179\\
EB&0.84&0.84&0.84&37\\
EW&0.96&0.91&0.91&93\\
GCAS&0.90&0.90&0.90&29\\
HADS&1.00&0.95&0.97&19\\
ROT&0.94&1.00&0.96&234\\
RRab&0.99&0.99&0.99&96\\
% RRc&0.95&0.97&0.96&108\\
% RRd&1.00&0.67&0.80&3\\
RRcd&0.97&1.00& 0.99&36\\
% T2CEP&1.00&0.40&0.57&5\\
UV&1.00&0.40&0.57&15\\
YSO&0.92&0.92&0.92&13\\
accuracy&&&0.96&794\\
weighted avg&0.96&0.96&0.95&794\\
\enddata
\end{deluxetable*}

\begin{figure*}
\centering
\begin{minipage}{185mm}
  \includegraphics[width=185mm]{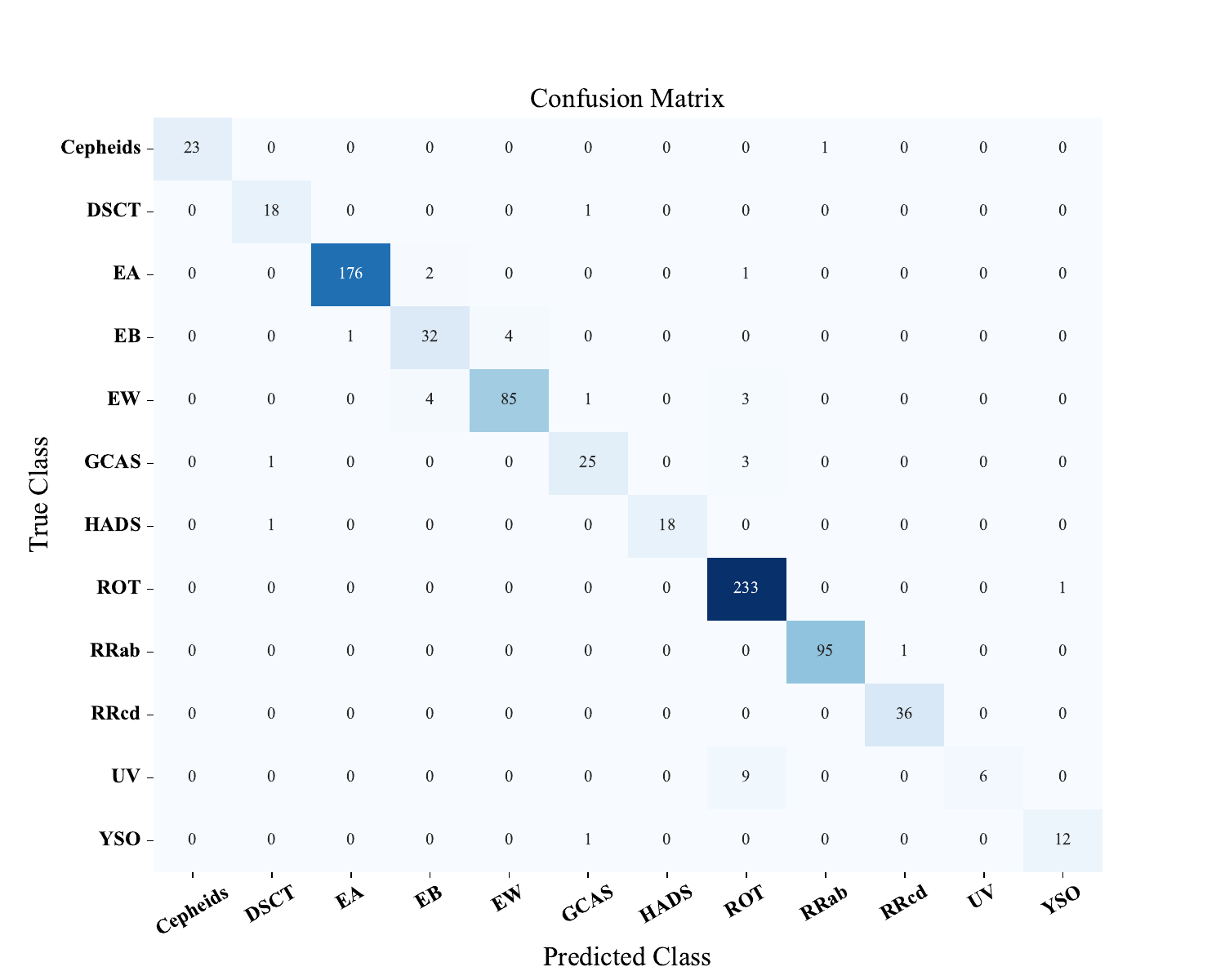}
\caption{{Confusion matrix of the classifier.}}
\label{Figure3}
  \end{minipage}
\end{figure*}

\section{Classification} \label{sec:Classification for Periodic Variable stars}
In order to classify periodic variable stars, commonly used features include the period, physical parameters, shape parameters and statistical parameters of the LC. Initially, we considered 27 parameters and we carefully analyzed the importance and independence of each parameter, and finally retained 19 physical parameters with high importance. A description of these parameters is listed in Table \ref{Table2}. We also indicated in Table 2 which parameters were estimated by the LC. We checked the distribution of each parameter to ensure the validity of the features, for example, we converted $\phi_{21}$ from $-2\pi\sim4\pi$ to between $\pi$ and 3$\pi$. 

Physical parameters include Gaia parallaxes, parallax uncertainties, $(B_P-R_P)_0$ and absolute Wesenheit magnitudes, and the relevant data were obtained by performing a 1-arcsec crossmatch to Gaia DR3 \citep{2022A&A...667A.148G,674A...1G}. About 1\% of objects are contaminated by other sources within 1 arcsec. Although we determined uniqueness by the help of TIC catalog, whether the Gaia parameters are affected by the contamination requires further examination based on the needs of the research. The intrinsic color was estimated by $(B_P-R_P)_0=G_{BP}-G_{RP}-E(B_P-R_P)$ and $E(B_P-R_P)$ was determined by the extinction value $A_G$ and the extinction law $A_G/E(B_P-R_P)=1.89$ from \citep{2019ApJ...877..116W}.  The absolute Wesenheit magnitudes were determined by $G-1.89\times E(B_{P}-R_{P})-5\log(100\times D)$. $D$ stands for distance (kpc) and is the inverse of the corrected Gaia parallaxes. The use of Wesenheit magnitude can avoid the extinction in the Galactic plane on the analysis of color--magnitude diagram and period--luminosity diagram. We also obtained $W_{1}-W_{3}$, $W_{1}-W_{4}$, and $M_{W_{1}}$ from WISE \citep{2010AJ....140.1868W}, which are parameters that can distinguish variable stars with circumstellar disks. $Q_{31}$ and Amp. are parameters that show the variation amplitude of LCs. The $Q_{31}$ value is the difference between the third quartile ($Q_{3}$) and the first quartile ($Q_{1}$) of a raw light curve, where $Q_{1}$ separates the lowest 25$\%$ from the highest 75$\%$ of the data, and $Q_{3}$ separates the lowest 75$\%$ from the highest 25$\%$. Amp represents the normalized peak-to-peak amplitude derived from the eighth-order Fourier fitting of the LCs. W, K, Std, $\gamma 2$, $\gamma 1$, $\phi_{21}$, $\phi_{31}$, $R_{21}$ and $R_{31}$ are parameters indicating the detailed shapes of the LCs. W refers to the Shapiro-Wilk statistic, used to test the null hypothesis that samples from the LC measurements follow a normal distribution (Shapiro $\&$ Wilk, 1965). The Stetson K parameter is calculated using a single-band LC by 
\begin{eqnarray}
K = \frac{1}{\sqrt{N}} \frac{\sum_{i=1}^{N}|\delta(i)|} {\sqrt{\sum_{i=1}^{N}\delta(i)^{2}}},
\label{eq:StetsonK}
\end{eqnarray}
where $N$ is the total number of measurements, $i$ is an index for each measurement, and $\delta(i)$ is the normalized residual for the i-th observation. `Std' is the standard deviation of the LC. $\gamma_{2}$ is the kurtosis of the LC and measures the tailedness of the distribution, which is defined by 
\begin{eqnarray}
\gamma_{2} = \frac{\frac{1}{n}\sum_{i=1}^{N}(x_{i}-\bar{x})^{4}}{(\frac{1}{n}\sum_{i=1}^{N}(x_{i}-\bar{x})^{2})^{2}} - 3.
\label{eq:kurtosis}
\end{eqnarray}
$\gamma_{1}$ is the skewness of the LC and measures the asymmetry, which is defined by 
\begin{eqnarray}
\gamma_{1} = \frac{\frac{1}{n}\sum_{i=1}^{N}(x_{i}-\bar{x})^{3}}{(\frac{1}{n}\sum_{i=1}^{N}(x_{i}-\bar{x})^{2})^{3}/2}.
\label{eq:skewness}
\end{eqnarray}
$\phi_{k1}$ and $R_{k1}$ can distinguish different shapes of light
curves, where $\phi_{k1}$ and $R_{k1}$ denote the phase difference and the amplitude ratio of the different components obtained from the Fourier fit.

%Parallaxes were calculated by parallaxes from Gaia DR3 minus zero-point correction from Gaia DR3.

\subsection{Training a classification model}\label{sec:Search for Periodic Variable stars}
 We cross-matched our TESS periodic variable stars with the ASAS-SN Catalog of Variable Stars in 1 arcsec to get a pre-classified set \citep{2023MNRAS.519.5271C}. We found the numbers of RR Lyrae stars, Cepheids, and eclipsing binaries in the pre-classified set are small and need to be supplemented. So we supplemented RR Lyrae stars and Cepheids from Gaia DR3 \citep{674A...1G}. 106 RRab, 29 RRc and 1 RRd were added to the pre-classified set, increasing the number of RRab, RRc and RRd from 185, 76 and 2 to 291, 105, and 3. 23 DCEP and 1 T2CEP were added to the pre-classified set, increasing the number of DCEP and T2CEP from 40 and 4 to 63 and 5. We also cross-matched our periodic variable stars with the TESS eclipsing binary catalog from \cite{2022ApJS..258...16P}. 17 EA and 6 EW were added to the pre-classified set, increasing the number of EA and EW from 525 and 276 to 542 and 282. We performed a visual inspection of LCs of the pre-classified set and excluded some of the variable stars with bad periodicity. 4 SX Phoenicis variable stars (SX PHE) were obtained, but we put them in the type of HADS because their LCs are very similar to those of HADS. In addition, the classifier trained with SX PHE as a separate type did not find new SX PHEs in the prediction set.

The final pre-classified set includes 2,405 periodic variable stars to train the classifier and test its accuracy. The pre-classified set includes 12 types: Cepheids (Cepheids include fundamental-mode classical Cepheids (DCEP), first-overtone classical Cepheids (DCEPS), and Type-II Cepheids (T2CEP).), DSCT, EA, EB, EW, GCAS, HADS, ROT, RRab, RRcd, UV and YSO. We used 2/3 of the pre-classified data as the training set and 1/3 of the pre-classified data as the testing set. Table \ref{Table3} shows the full name and number of each type. We used a random forest classifier as the machine learning model, which is based on an ensemble of decision trees \citep{1993cpml.book.....Q,2001MachL..45....5B}. We used grid search to find the optimal combination of parameters and used the testing set to get the accuracy of the classifier. We obtain the importance of each feature and show them in Figure \ref{Figure2}. During the construction of each decision tree, whenever a feature is used to split a node, the algorithm calculates its overall contribution to the model. The higher the contribution, the more important the feature is in improving classification performance. In Table \ref{Table4}, we show the precision, recall and the $F_1$-score for each variable star type, i.e.,
\begin{equation}
    precision={\rm \frac{True\;Positive}{True\;Positive + False\;Positive}}~,
\end{equation}
\begin{equation}
    recall={\rm \frac{True\;Positive}{True\;Positive + False\;Negative}}~,
\end{equation}
\begin{equation}
    F_1=2\times \frac{precision\times recall}{precision + recall}~,
\end{equation}
and $F_1$ ranges between 0 and 1. Physical parameters that are important for the classification of variable stars are broadly classified into the following three categories: periods, LCs parameters (parameters related to the amplitudes of LCs, such as $Q_{31}$ and amplitudes, and parameters related to the shapes of LCs, such as W, K, $\phi_{21}$, $\phi_{31}$, $\gamma 2$, $\gamma 1$, and two parameters related to amplitudes ratios: $R_{21}$ and $R_{31}$), and physical parameters. The important physical parameters are $M_{W_{G}}$, parallaxes, $M_{W_{1}}$, and $(B_{P}-R_{P})_{0}$. The important parameters for each type of variable stars are discussed in Section \ref{sec:Discussion of Periodic Variable stars}.

\begin{table*}[ht!]
\vspace{-0.0in}
\begin{center}
\caption{\label{Table5}TESS Periodic Variable Star Catalog\tablenotemark{a}.}
\vspace{0.15in}
\begin{tabular}{lcccccccccccl}
\hline
\hline
TIC  & R.A. (J2000)    & Dec. (J2000)   & Period   & $(B_{P}-R_{P})_{0}$ &$M_{W_{G}}$    &...&Correct Classification Probability& Type  \\      
    & $^\circ$ & $^\circ$ & days    &    mag      &    mag          &  ...    &     &        \\
\hline     
       358& 218.79892 &-29.34861 &  4.390105& 0.5268  & 2.1325&...& 0.43  &EA   \\  
     10475& 219.06313 &-24.39062 &  4.830116& 0.7606  & 1.7823&...& 0.63  &ROT  \\  
     11182& 219.15315 &-24.76302 &  0.973735& 2.8996  &10.3371&...& 0.70  &ROT  \\  
     11582& 219.15724 &-25.52329 &  4.688072& 1.7452  &-2.6580&...& 0.49  &GCAS \\  
     12190& 219.16503 &-26.57896 &  0.099892& 0.4552  & 1.7777&...& 0.39  &GCAS \\  
     15093& 219.25620 &-28.72182 &  0.411680& 0.5417  & 2.7349&...& 0.37  &GCAS \\  
     16056& 219.27416 &-27.12844 &  2.875456& 0.7764  & 4.6822&...& 0.71  &ROT  \\  
     22370& 219.55623 &-28.61946 &  3.333288& 0.9587  & 5.5042&...& 0.94  &ROT  \\  
     22776& 219.60211 &-28.00178 &  2.303143& 0.7706  & 4.4063&...& 0.62  &ROT  \\  
     29069& 219.76181 &-29.63770 & 11.863676& 2.8534  &-0.9605&...& 0.43  &GCAS \\  
     29465& 219.84581 &-29.67732 & 16.368252& 2.6326  &-1.2611&...& 0.49  &GCAS \\  
     31122& 219.86874 &-27.10725 &  0.585084& 0.7834  & 3.9944&...& 0.49  &ROT  \\  
     33905& 220.00682 &-25.24255 &  2.727490& 2.5667  &10.0794&...& 0.70  &ROT  \\  
     34884& 219.91787 &-26.66621 &  1.384755& 0.6455  & 3.7708&...& 0.47  &ROT  \\  
     34900& 219.94593 &-26.69806 &  2.407725& 1.3240  & 6.4785&...& 0.89  &ROT  \\  
     34920& 219.92064 &-26.72365 &  2.395659& 0.6325  & 4.0443&...& 0.39  &ROT  \\  
     40238& 220.11161 &-25.76141 &  0.980546& 0.4622  & 2.4992&...& 0.45  &ROT  \\  
     42399& 220.22626 &-25.54776 &  0.951238& 0.4896  & 2.7684&...& 0.41  &ROT  \\  
     46937& 220.40486 &-27.85877 &  0.194518& 2.1754  & 8.8228&...& 0.53  &ROT  \\  
     48324& 220.35386 &-25.83843 &  0.611392& 0.5590  & 2.5227&...& 0.44  &GCAS \\  
     51431& 220.52218 &-26.77729 &  5.973201& 1.7316  & 7.6098&...& 0.98  &ROT  \\  
     52357& 220.48820 &-28.06576 &  0.548154& 1.5612  &-0.3182&...& 0.24  &EW   \\  
     55344& 220.60762 &-27.78729 &  9.543846& 1.9877  &-1.5155&...& 0.51  &GCAS \\  
     60583& 220.71778 &-28.25488 &  1.004029& 0.4286  & 2.6118&...& 0.49  &ROT  \\  
     60627& 220.79599 &-28.30994 &  4.734238& 0.9100  & 5.0722&...& 0.79  &ROT  \\  
     62545& 220.85235 &-29.08589 &  2.162700& 0.6723  & 3.8171&...& 0.55  &ROT  \\  
     65628& 220.87720 &-24.45908 &  0.399619&-0.0200  & 0.1793&...& 0.68  &GCAS \\  
     65919& 220.95218 &-24.01521 &  2.903746& 0.8398  & 4.7504&...& 0.76  &ROT  \\  
     66094& 221.07383 &-24.12870 &  4.809058& 1.3429  &-1.1269&...& 0.43  &GCAS \\  
     67228& 221.08765 &-25.89498 &  5.159653& 0.9717  & 4.6387&...& 0.67  &ROT  \\  
     68069& 221.06245 &-27.29045 &  0.379272& 3.1095  &11.6656&...& 0.74  &ROT  \\  
     70111& 221.20410 &-29.69859 &  0.521713& 2.4877  & 9.4565&...& 0.71  &ROT  \\  
     70654& 221.11463 &-29.01342 &  0.064387& 0.5644  & 2.5213&...& 0.32  &DSCT \\  
     82066& 221.50034 &-25.44318 &  5.054140& 0.4055  & 2.5938&...& 0.39  &ROT  \\  
     90907& 221.75130 &-25.46473 &  4.016067& 0.6494  & 3.8914&...& 0.50  &ROT  \\  
     92193& 221.88302 &-24.20143 &  0.230209&-0.2168  &-0.9188&...& 0.76  &GCAS \\  
     93155& 221.84398 &-25.62427 &  2.200173&-0.0059  &-0.5321&...& 0.80  &GCAS \\  
     96608& 221.90784 &-29.74424 &  1.590696& 0.6218  & 1.5152&...& 0.50  &ROT  \\  
    101462& 222.05765 &-24.61321 & 11.264432& 0.8746  & 5.1828&...& 0.74  &ROT  \\  
    102723& 222.13313 &-26.47361 &  4.011592& 2.5179  & 9.7719&...& 0.82  &ROT  \\  

   ...     & ...      &...       &  ...       & ...      &  ...     & ...   &  ...    &...   \\ 
\hline
\end{tabular}
\tablenotetext{a}{The entire table is available in the online journal;
  40 lines are shown here for guidance regarding its form and
  content.}
\end{center}
\end{table*}

\begin{figure*}
\centering
\begin{minipage}{185mm}
  \includegraphics[width=185mm]{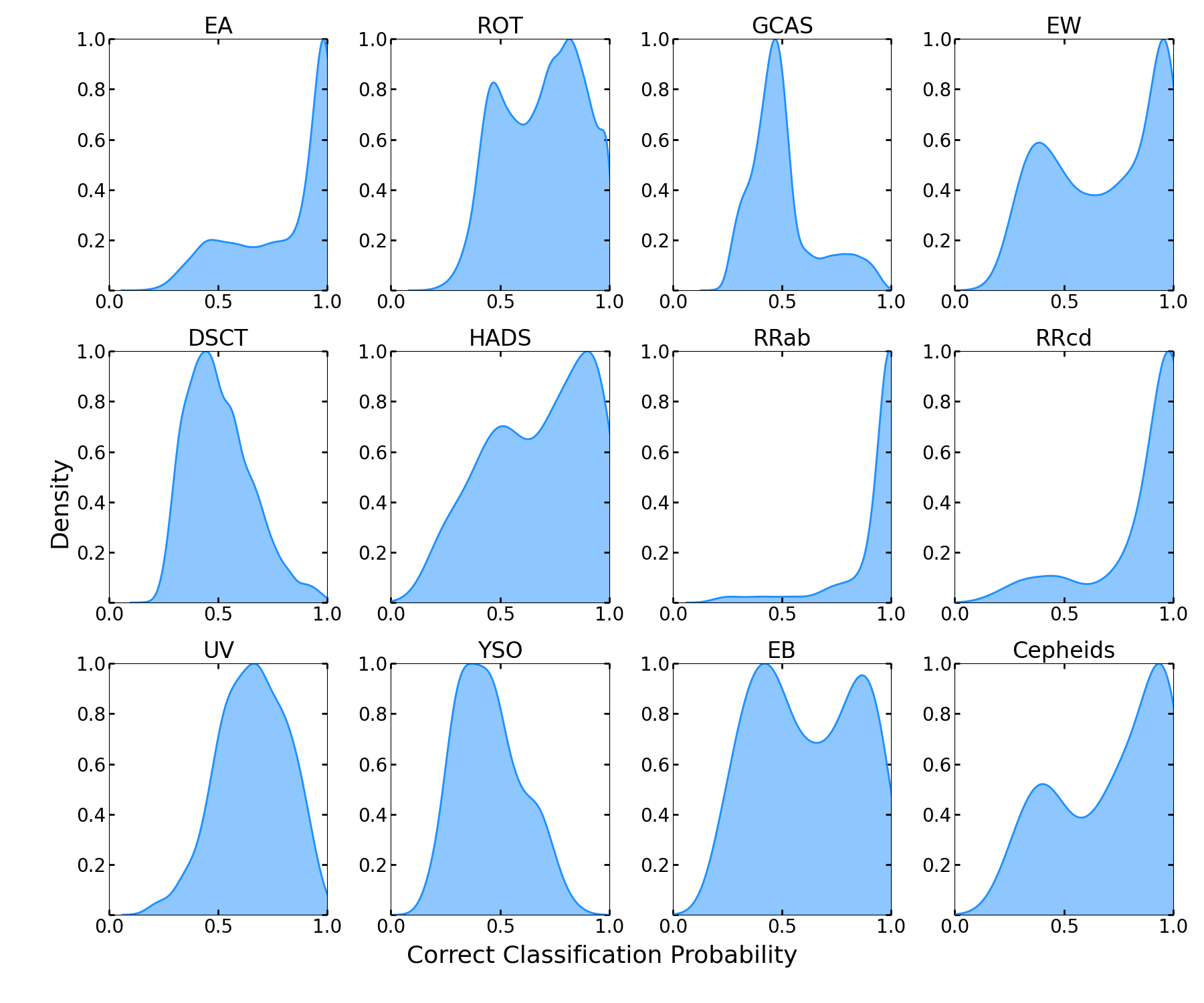}
\caption{{Distributions of the correct classification probability of 12 different variable stars.}}
\label{Figure4}
  \end{minipage}
\end{figure*}

\begin{figure*}
\centering
\begin{minipage}{185mm}
  \includegraphics[width=185mm]{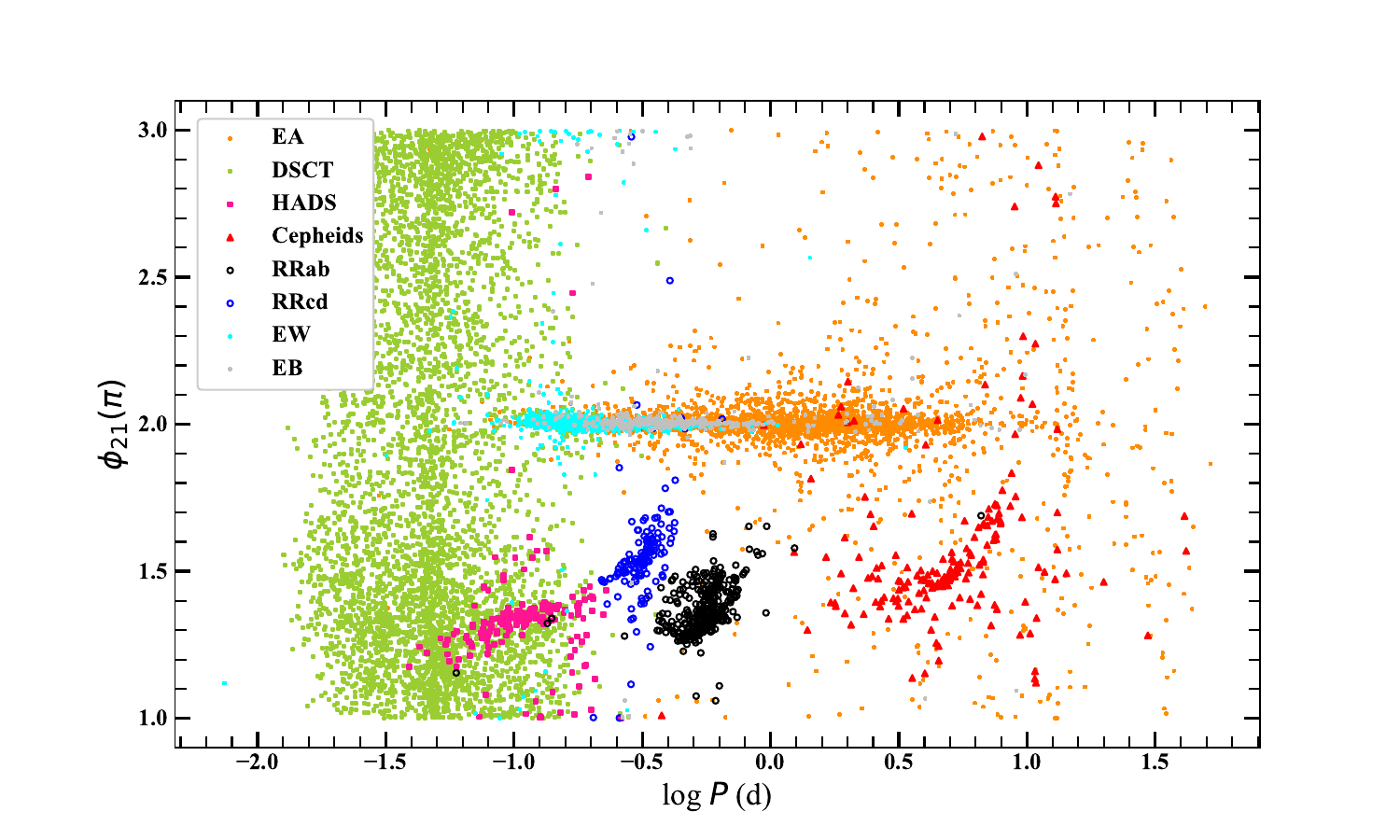}
\caption{$\phi_{21} (\pi)$ vs. $\log P$ diagram for TESS variable stars with amplitudes larger than 0.006 mag. From short to long periods, there are $\delta$ Scuti stars (DSCT; yellowgreen), high-amplitude $\delta$ Scuti stars (HADS; pink), EW-type eclipsing binaries (EW; cyan), EB-type eclipsing binaries (EB; silver), EA-type eclipsing binaries (EA; orange), fundamental RR Lyrae (RRab; black circles), first-overtone and double-mode RR Lyrae (RRcd; blue circles), Cepheids (red triangles).}
\label{Figure5}
  \end{minipage}
\end{figure*}

\begin{deluxetable*}{ccc}
\tablecaption{{ All and Newly Classified TESS Variable stars.}\label{Table6}}
\tablehead{\colhead{Type}& \colhead{Total}& \colhead{New (Fraction)}}
\startdata
Cepheids&184&13 (7.1$\%$)\\
% DCEP&164&12 (7.3$\%$)\\
DSCT&4887&4620 (94.5$\%$)\\
EA&3938&1306 (33.2$\%$)\\
EB&404&58 (14.4$\%$)\\
EW&1301&292 (22.4$\%$)\\
GCAS&23641&22166 (93.8$\%$)\\
HADS&184&40 (21.7$\%$)\\
ROT&36740&33900 (92.3$\%$)\\
RRab&333&18 (5.4$\%$)\\
RRcd&139&9 (6.5$\%$)\\
% T2CEP&7&1 (14.3$\%$)\\
% UG&92&81 (88.0$\%$)\\
UV&477&441 (92.5$\%$)\\
YSO&277&242 (87.4$\%$)\\
\enddata
\end{deluxetable*}

\begin{figure*}
\centering
\begin{minipage}{185mm}
  \includegraphics[width=185mm]{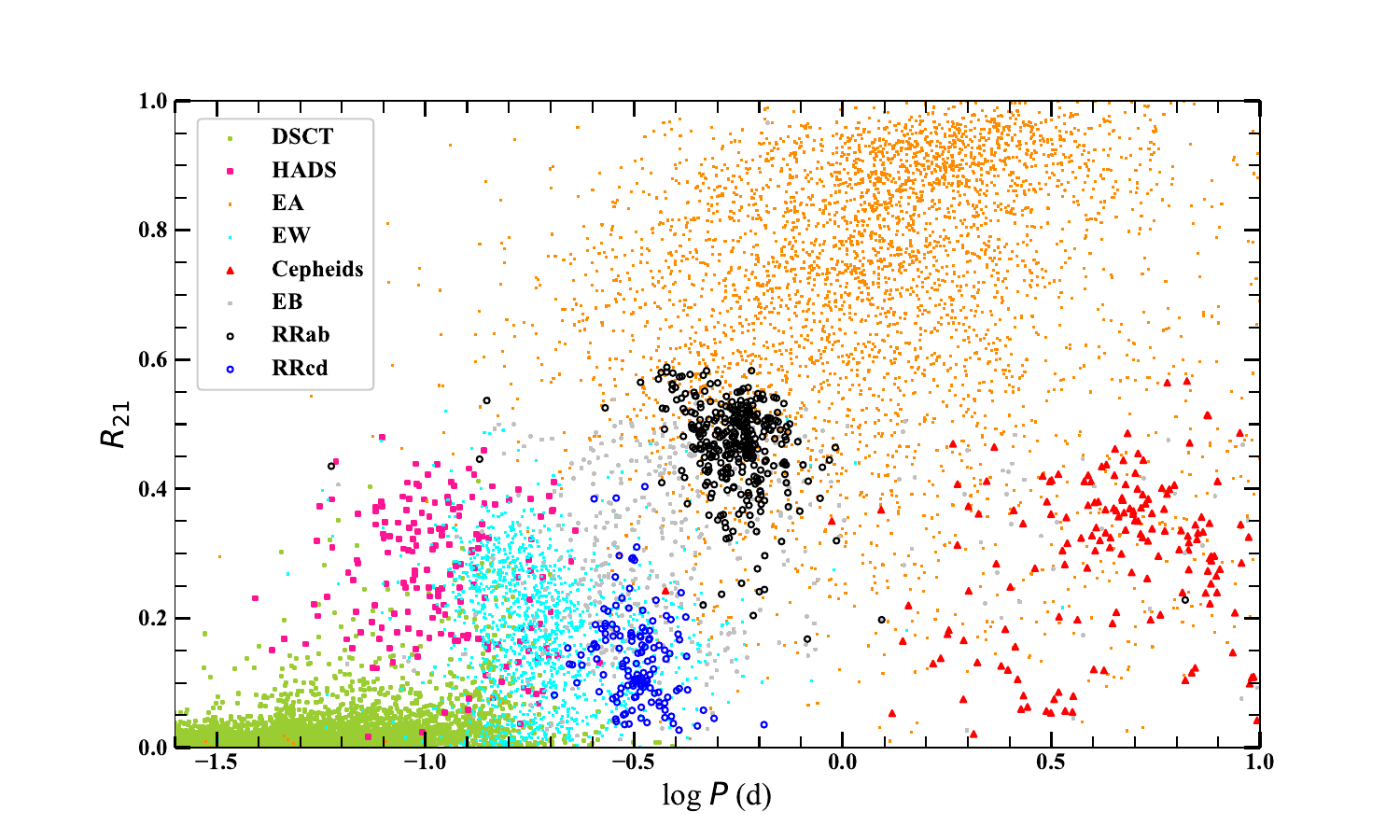}
\caption{$R_{21}$ vs. $\log P$ diagram for TESS variable stars. Symbols are as in Figure 4.}
\label{Figure6}
  \end{minipage}
\end{figure*}

\begin{figure*}
\centering
\begin{minipage}{185mm}
  \includegraphics[width=185mm]{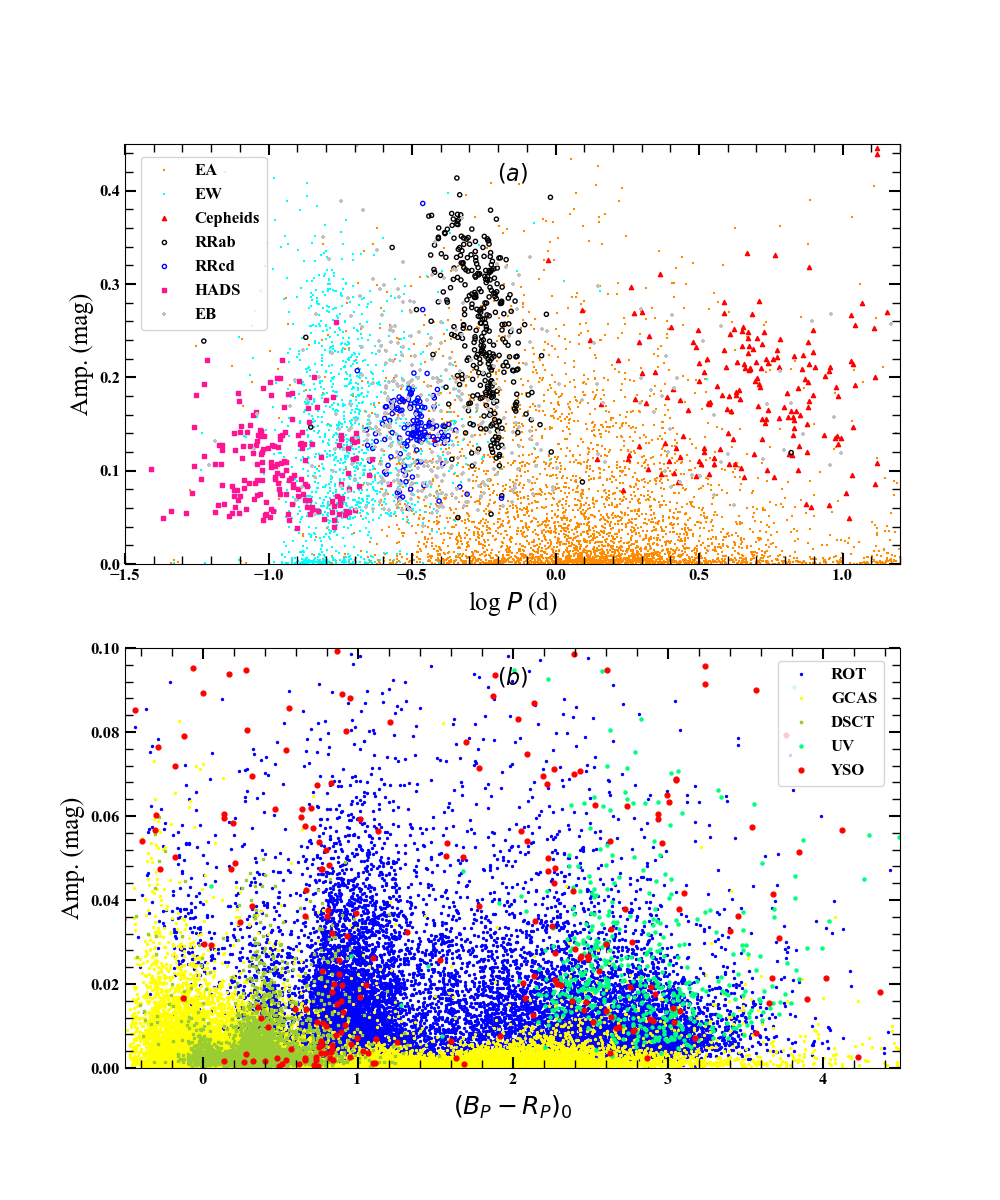}
\caption{Panel (a) shows Amp. (mag) vs. $\log P$ diagram for high-amplitude TESS variable stars. Panel (b) shows Amp. (mag) vs. $(B_{P}-R_{P})_{0}$ (mag) diagram for low-amplitude TESS variable stars. Symbols are as in Figure \ref{Figure4}. GCAS (yellow dots), ROT (blue solid circles), UV (springgreen dots), and YSO (red solid circles) are also included.}
\label{Figure7}
  \end{minipage}
\end{figure*}

\begin{figure*}
\centering
\begin{minipage}{185mm}
  \includegraphics[width=185mm]{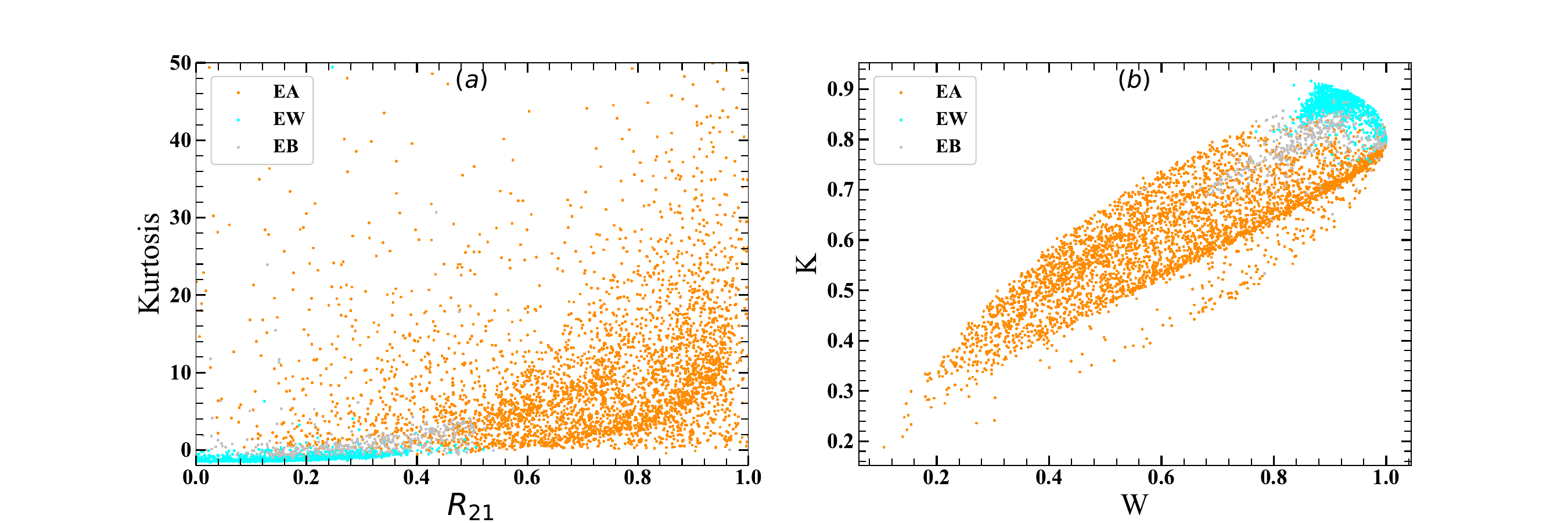}
\caption{The left panel shows the Kurtosis vs. $R_{21}$ diagram and the right panel shows the K vs. W diagram. Symbols are as in Figure \ref{Figure4}.}
\label{Figure8}
  \end{minipage}
\end{figure*}

\begin{figure*}
\centering
\begin{minipage}{185mm}
  \includegraphics[width=185mm]{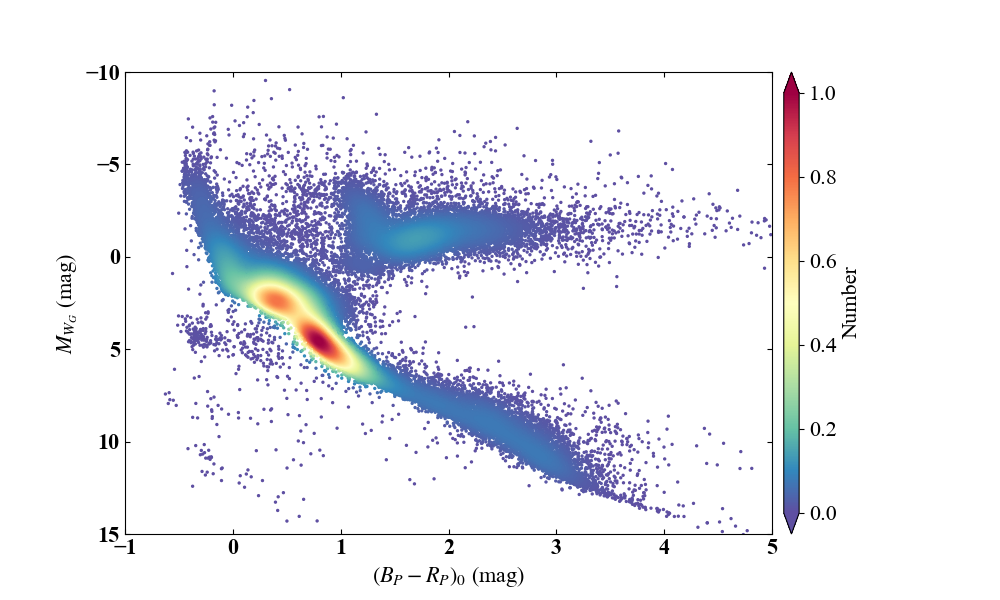}
\caption{The scattered density map showing $(B_{P}-R_{P})_{0}$ (mag) -- $M_{W_{G}}$ (mag) distribution. The number in the colorbar is the normalized number.}
\label{Figure9}
\end{minipage}
\end{figure*}

\begin{figure*}
\centering
\begin{minipage}{185mm}
  \includegraphics[width=185mm]{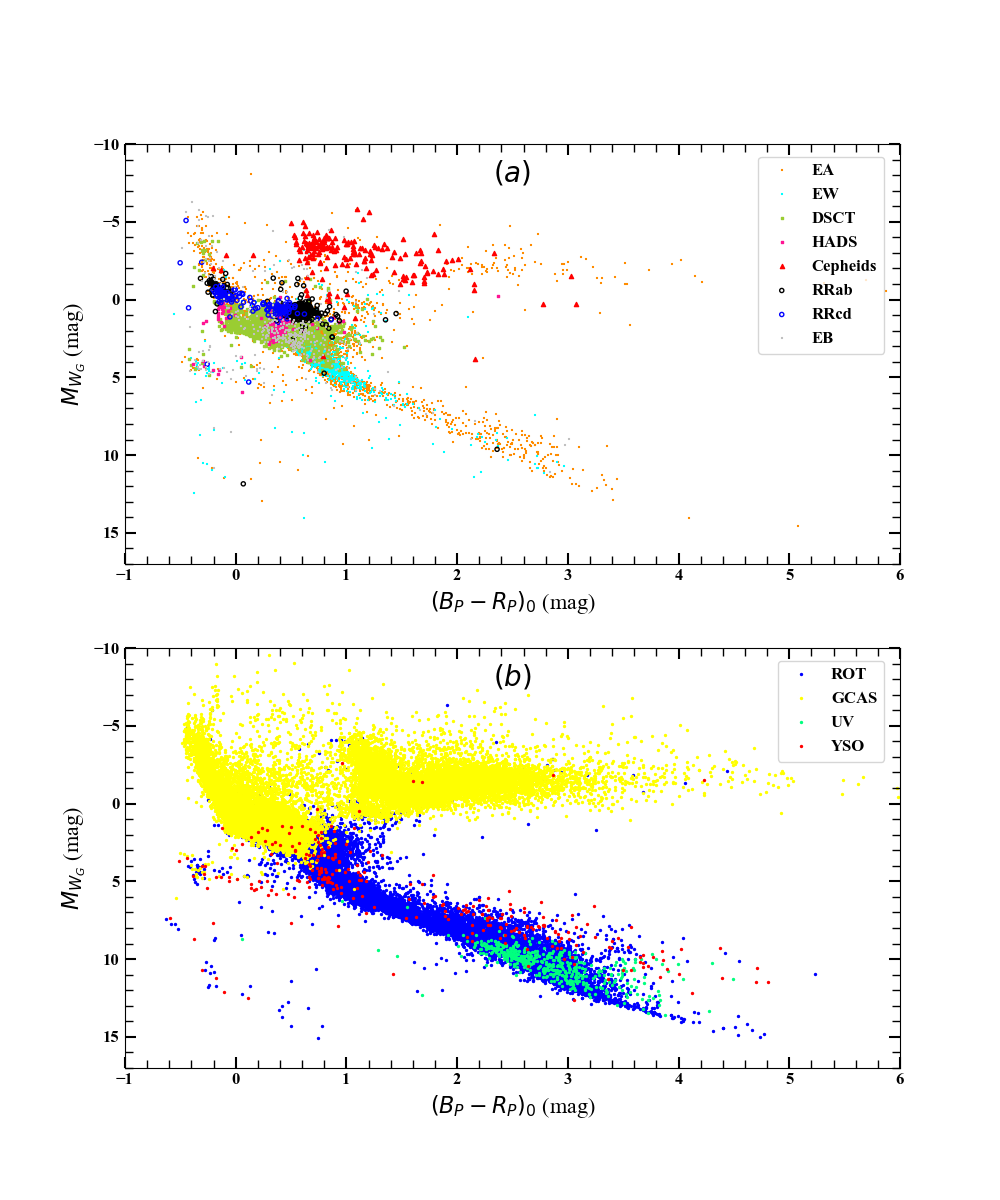}
\caption{The $(B_{P}-R_{P})_{0}$ (mag) -- $M_{W_{G}}$ (mag) diagram for all TESS variable stars. Symbols in the upper panel are as in Figure \ref{Figure4}. Symbols in the lower panel are as in Figure \ref{Figure6}.}
\label{Figure10}
\end{minipage}
\end{figure*}

\begin{figure*}
\centering
\begin{minipage}{185mm}
  \includegraphics[width=185mm]{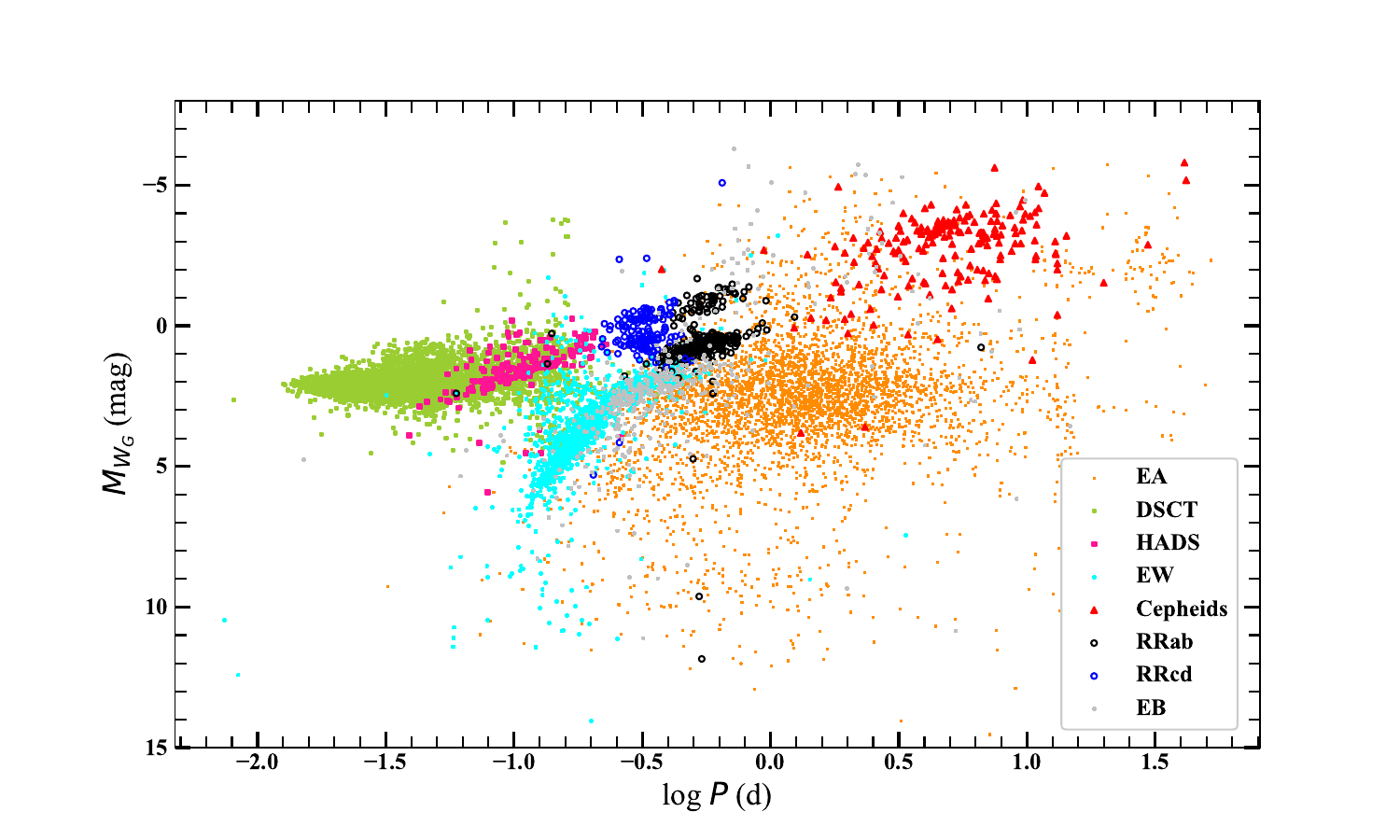}
\caption{$\log P$ -- $M_{W_{G}}$ (mag) diagram for TESS variable stars. Symbols are as in Figure \ref{Figure4}.}
\label{Figure11}
\end{minipage}
\end{figure*}

\begin{figure}[htbp]
\centering
\subfigure[]{
\begin{minipage}{185mm}
  \includegraphics[width=185mm]{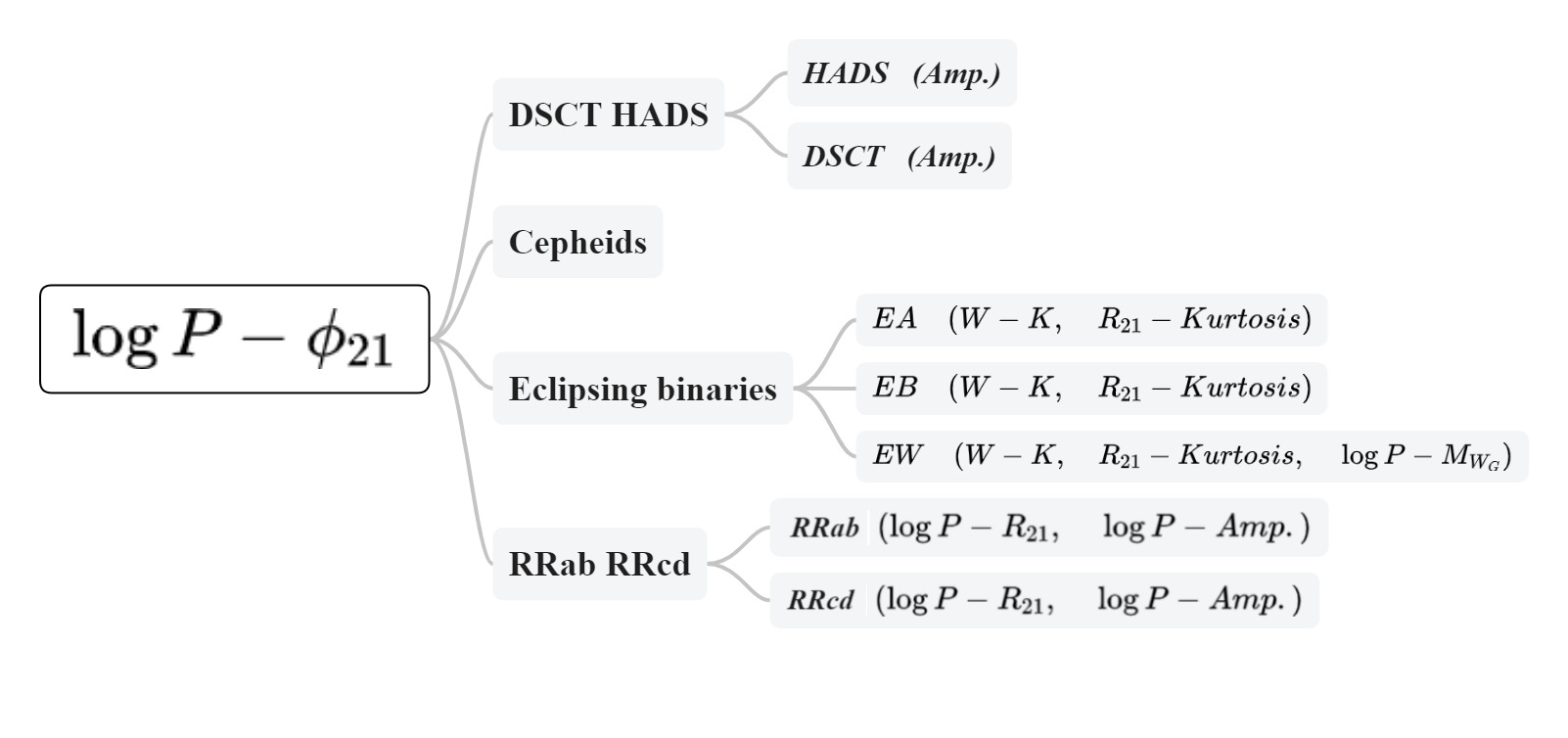}
\end{minipage}
}

\subfigure[]{
\begin{minipage}{150mm}
    \includegraphics[width=150mm]{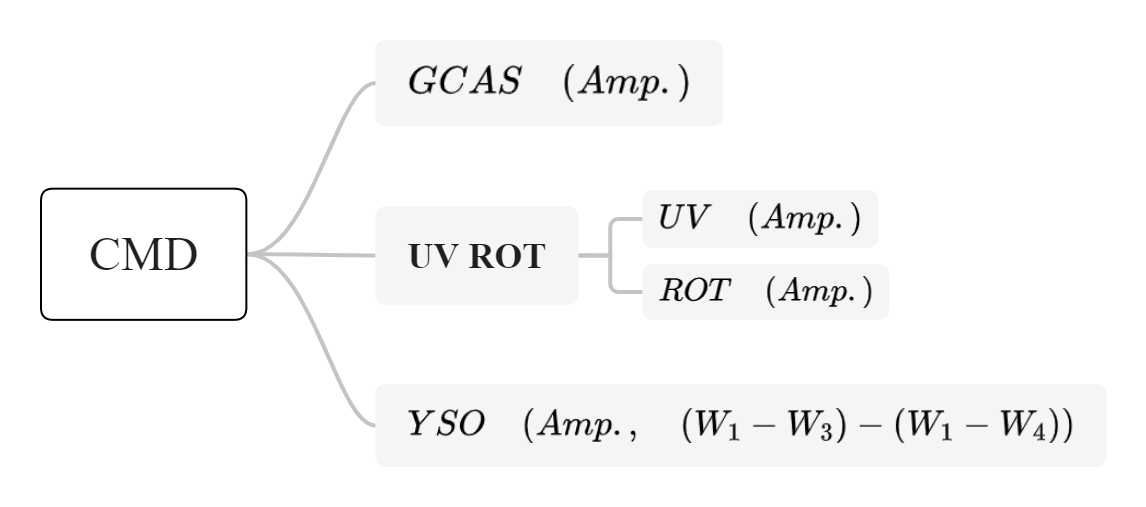}
\end{minipage}
}
\caption{Important parameters for variable star classification in machine learning. The upper panel includes pulsation stars and eclipsing binaries, and the lower panel includes GCAS, ROT, UV, and YSO.}
\label{Figure12}
\end{figure}

Table \ref{Table4} shows the classification performance of the classifier and Figure \ref{Figure3} shows the confusion matrix. As shown in Table \ref{Table4}, the weighted average of precision, recall and the $F_1$-score for all types reaches 0.96, which proves that the classifier successfully classifies the training set. From Figure \ref{Figure3}, we can see the main contamination of each type of variable star. When only pulsation variable stars and eclipsing binaries are counted, the weighted average $F_1$-score is up to 0.96. The $F_1$-score of Cepheids is 0.98, and a very small number of Cepheids have been misclassified as RR Lyrae. This is because their LCs are similar to those of RR Lyrae while RR Lyrae are more numerous in the training set. The $F_1$-score of EA is 0.99 and the $F_1$-score of EW is 0.91. From Figure \ref{Figure3}, we find that the main contaminants of eclipsing binaries are almost the other sub-type eclipsing binaries. $F_1$-scores of both DSCT (0.95) and HADS (0.97) are higher than 0.90, suggesting that only a small number of DSCT and HADS are misclassified. RRab and RRcd are well classified by the classifier. The $F_1$-score of RRab is up to 0.99, and that of RRcd is 0.99. The $F_1$-score of ROT is up to 0.96, and $F_1$-scores of GCAS and YSO are 0.90 and 0.92, respectively. $F_1$-score of UV (0.57) is lower, because they have fewer features in LCs.

In this section, we used a pre-classified data consisting of a sample of 2405 variable stars to train and test the random forest classifier. The periodic variable stars were classified into 12 sub-types, including Cepheids, $\delta$ Scuti variable stars (DSCT, and HADS), eclipsing binaries (EA, EB, and EW), RR Lyrae (RRab, and RRcd), GCAS, ROT, UV, and YSO. The classification performance of the random forest classifier is well and the weighted average of precision and recall in the training set reach 0.96.

\begin{deluxetable*}{ccccc}
\tablecaption{{Variable Star Purity Comparison between TESS and Other Catalogs.}\label{Table7}}
\tablehead{\colhead{Type}&\colhead{Total}& \colhead{Gaia DR3}& \colhead{ZTF DR2}& \colhead{Total Sample}}
\startdata
Cepheids&185&94.0$\%$&100.0$\%$&94.4$\%$\\
% DCEP&97.3$\%$&100.0$\%$&97.6$\%$\\
% DCEPS&100.0$\%$&-&100.0$\%$\\
DSCT&4982&96.1$\%$&100.0$\%$&96.2$\%$\\
EA&3935&94.4$\%$&90.1$\%$&94.2$\%$\\
EB&392&99.4$\%$&-&99.4$\%$\\
EW&1260&99.0$\%$&92.1$\%$&98.0$\%$\\
HADS&215&97.8$\%$&100.0$\%$&98.1$\%$\\
ROT&35805&82.3$\%$&89.3$\%$&83.3$\%$\\
RRab&340&98.7$\%$&98.7&98.7$\%$\\
RRcd&149&97.6$\%$&90.0$\%$&96.6$\%$\\
% RRd&100.0$\%$&-&100.0$\%$\\
% T2CEP&100.0$\%$&-&100.0$\%$\\
\enddata
\end{deluxetable*}

\begin{deluxetable*}{ccccc}
\tablecaption{{ Variable Star Purity Comparison between TESS and Other Catalogs for objects with correct classification probability greater than 0.5.}\label{Table8}}
\tablehead{\colhead{Type}&\colhead{Total}& \colhead{Gaia DR3}& \colhead{ZTF DR2}& \colhead{Total Sample}}
\startdata
Cepheids&133&99.2$\%$&100.0$\%$&99.3$\%$\\
% DCEP&97.3$\%$&100.0$\%$&97.6$\%$\\
% DCEPS&100.0$\%$&-&100.0$\%$\\
DSCT&2181&99.1$\%$&100.0$\%$&99.1$\%$\\
EA&3373&95.5$\%$&92.4$\%$&95.4$\%$\\
EB&240&99.6$\%$&-&99.6$\%$\\
EW&885&99.7$\%$&99.2$\%$&99.6$\%$\\
HADS&133&98.6$\%$&100.0$\%$&98.9$\%$\\
ROT&28852&91.9$\%$&93.3$\%$&92.1$\%$\\
RRab&318&100.0$\%$&100.0&100.0$\%$\\
RRcd&123&100.0$\%$&100.0$\%$&100.0$\%$\\
% RRd&100.0$\%$&-&100.0$\%$\\
% T2CEP&100.0$\%$&-&100.0$\%$\\
\enddata
\end{deluxetable*}

\begin{deluxetable*}{ccccc}
\tablecaption{{Variable Star Purity Comparison between TESS and Other Catalogs of the 757 objects with different periods.}\label{Table9}}
\tablehead{\colhead{Type}&\colhead{Total}& \colhead{Gaia DR3}& \colhead{ZTF DR2}& \colhead{Total Sample}}
\startdata
Cepheids&21&100.0$\%$&94.1$\%$&94.7$\%$\\
% DCEP&97.3$\%$&100.0$\%$&97.6$\%$\\
% DCEPS&100.0$\%$&-&100.0$\%$\\
EA&300&25.0$\%$&80.8$\%$&77.9$\%$\\
EB&7&-&100.0$\%$&100.0$\%$\\
EW&4&-&100.0$\%$&100.0$\%$\\
ROT&382&86.4$\%$&75.0$\%$&83.3$\%$\\
RRab&4&-&100.0$\%$&100.0$\%$\\
RRcd&5&-&100.0$\%$&100.0$\%$\\
% RRd&100.0$\%$&-&100.0$\%$\\
% T2CEP&100.0$\%$&-&100.0$\%$\\
\enddata
\end{deluxetable*}

\subsection{The classifier on the TESS datasets}\label{sec:The classifier on the TESS datasets}
Based on the random forest classifier, we obtain the predicted types of 70,100 unclassified variable stars and list the number of each type in Table \ref{Table3}. To help distinguish between the easier and harder to classify objects within each type, we introduced a parameter correct classification probability, as shown in Table \ref{Table5}. For each variable star, the classifier gives the probability of 12 types that sums to 1. The type predicted by the classifier are based on the highest probability. We used the highest probability predicted by the model as the correct classification probability. The type of objects with correct classification probability above 0.5 is considered to be more reliable. 25,734 objects have a correct classification probability less than 0.5. 14,380 ($\sim$ 55.9$\%$) and 7.201 ($\sim$ 28.0$\%$) objects with correct classification probability less than 0.5 belong to GCAS and ROT, respectively. Nevertheless, there are periodic variable stars that have more than one type at the same time, which results in a lower correct classification probability.

Figure \ref{Figure4} shows the kernel density estimation (KDE) plots of the correct classification probability for 12 different types of variable stars, including Cepheids, DSCT, EA, EB, EW, GCAS, HADS, ROT, RRab, RRcd, UV, and YSO. The x-axis represents the correct classification probability and the y-axis represents the density. For EA and RR Lyrae, their correct classification probabilities are mostly above 0.8 which means that they are easy to classify. For Cepheids, EW, and HADS, their distributions appear to have 2 peaks. For Cepheids, EW, and HADS with correct classification probability less than 0.5, the type with second highest classification probability is RRab, other types of eclipsing binaries and DSCT, respectively. This suggests that these variable stars are not too difficult to classify, or at least very accurate for the main-type classification. For DSCT, GCAS, ROT, UV, and YSO, the classification is more difficult due to the small amplitude and fewer features of the parameters. The physical parameters are very important for the classification of small-amplitude variable stars. In Figures \ref{Figure5} to \ref{Figure11}, we show how different physical parameters are used for different variable star classifications. The physical interpretations are discussed in Section \ref{sec:Discussion of Periodic Variable stars}. Figure \ref{Figure12}a and Figure \ref{Figure12}b summarize our classification criteria for different variable stars.

Figure \ref{Figure5} shows $\log P$ -- $\phi_{21}$ distribution of variable stars. The distributions of each type of variable stars are similar to the results from \cite{2020ApJS..249...18C}. The $\phi_{21}$ of eclipsing binaries is around $2\pi$, distinctly different from pulsation stars. With the period increases, pulsation stars can be divided into DSCT and HADS ($P \textless 0.2$ d), RRcd (0.2 d $\textless P \textless$ 0.4 d), RRab (0.4 d $\textless P \textless$ 1.0 d), and Cepheids (1.0 d $\textless P$). The $\log P$ -- $R_{21}$ distribution is shown in Figure \ref{Figure6}. The majority of variable stars have $R_{21}$ values between 0 and 1.

Figure \ref{Figure7}a shows the $\log P$ -- Amp. distribution of high-amplitude variable stars, while Figure \ref{Figure7}b shows the $(B_{P}-R_{P})_{0}$ -- Amp. distribution of low-amplitude variable stars. Amp. refers to the peak-to-peak amplitude determined from the eighth-order Fourier fit to the LCs. We separate the variable stars with low amplitudes (DSCT, ROT, GCAS, UV and YSO) and high amplitudes (HADS, eclipsing binaries, RR Lyrae and Cepheids). We find that the amplitudes of high-amplitude variable stars are mainly in the range of 0 to 0.4 mag. The amplitudes of HADS are almost larger than 0.05 mag (Figure \ref{Figure7}a), while the amplitudes of DSCT are almost lower than 0.05 mag (Figure \ref{Figure7}b).

Figure \ref{Figure8}a shows the $R_{21}$ -- Kurtosis diagram and Figure \ref{Figure8}b shows the W -- K diagram. Since only the eclipsing binaries have clear features in these figures, we do not show other variable stars to ensure the cleanliness and clarity of the figure. Figure \ref{Figure9} is a scatter density plot showing the distribution of $(B_{P}-R_{P})_{0}$ -- $M_{W_{G}}$. Numbers are the normalized values. Figure \ref{Figure10} shows the $(B_{P}-R_{P})_{0}$ -- $M_{W_{G}}$ diagram (CMD). Variable stars are located on the instability strip, the main sequence, the giant branch, and the white dwarf sequence. Figure \ref{Figure11} shows the $\log P$ -- $M_{W_{G}}$ diagram. We show all pulsation stars and eclipsing binaries in this figure.

\section{The Periodic Variable star Catalog} \label{sec:The Periodic Variable stars Catalog}
Our periodic variable star catalog of TESS, including 72,505 periodic variables, is listed in Table \ref{Table5}. It includes the source ID, position (J2000 R.A. and decl.), period, LC parameters, physical parameters, the correct classification probability and type of each variable star. The LC parameters include the amplitude, amplitude ratios $R_{21}$ and $R_{31}$, phase differences $\phi_{21}$ and $\phi_{31}$, Kurtosis, Skewness, $Q_{31}$, W, K, and Std. The physical parameters include the parallax and its error, $(B_{P}-R_{P})_{0}$ and $M_{W_{G}}$, $W_{1}-W_{3}$, $W_{1}-W_{4}$, and $M_{W_{1}}$. The full light curve images can be downloaded from \url{https://nadc.china-vo.org/res/r101482/}.

\begin{figure*}
\centering
\begin{minipage}{185mm}
  \includegraphics[width=185mm]{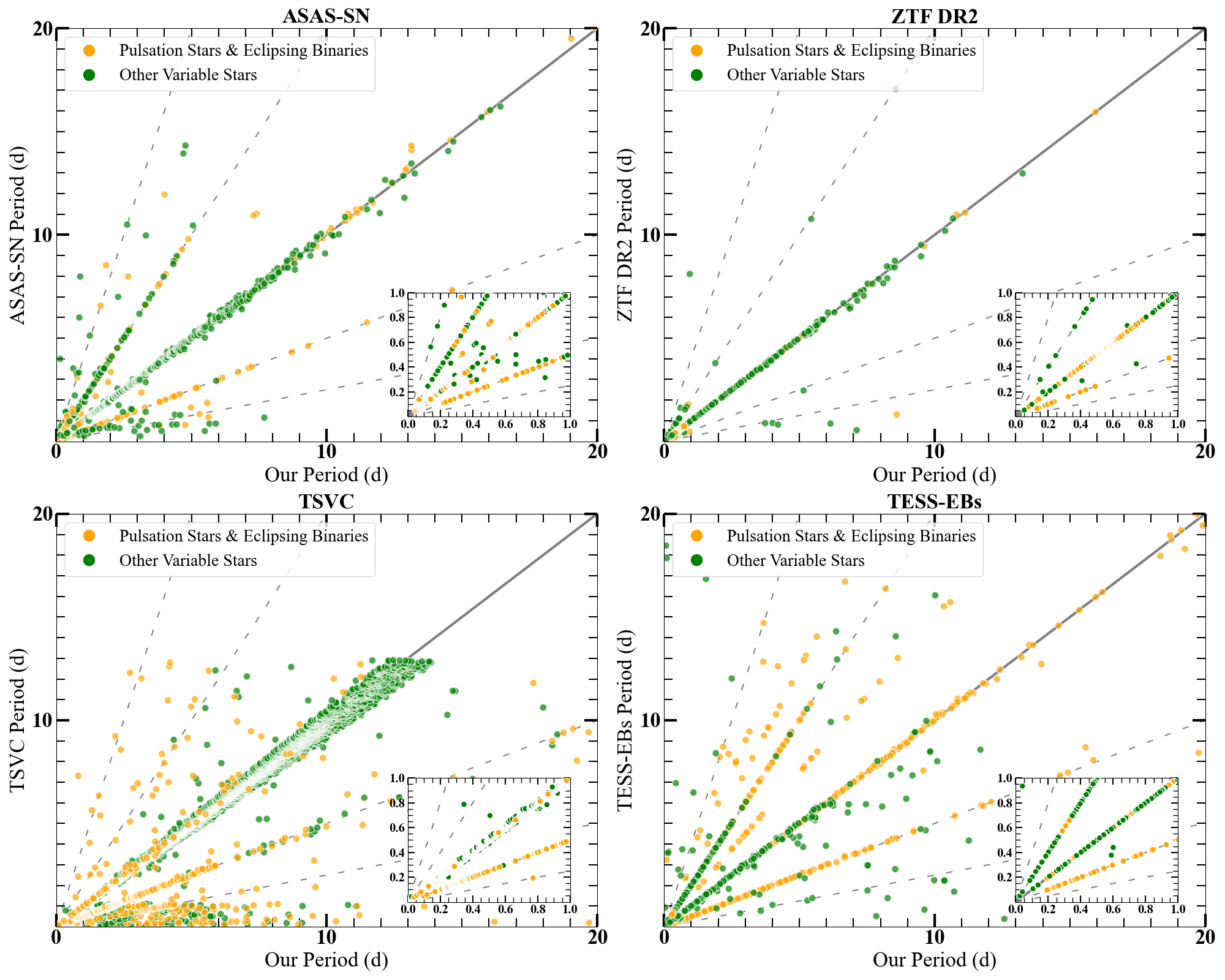}
\caption{ Comparison between our variability periods and the periods of four other variable star catalogs: ASAS-SN (top left), ZTF DR2 (top right), TSVC (bottom left), and TESS-EBs (bottom right). The solid gray line highlights the matching variability periods. The dashed gray lines show the lines where our periods are different from periods in other catalogs by factors of two or four. The bottom right corner of each panel shows zoomed-in view of the comparison characterized by $0<P<1$ days.}
\label{Figure13}
\end{minipage}
\end{figure*}

\subsection{New classified Variable stars} \label{sec:New classified Variable stars}
We cross-matched our periodic variable star catalog with Gaia DR3 \citep{2023A&A...674A..13E} and ZTF DR2 \citep{2020ApJS..249...18C} using an angular radius of 1 arcsec, to check the accuracy of our classification. Because the eclipsing binaries classification of TESS-EBs catalog has high accuracy, we also cross-matched our variable star catalog with TESS-EBs \citep{2022ApJS..258...16P} and found 2,945 matched variable stars. A total of 63,106 of the 72,505 variable stars in our catalog are newly classified. New variable stars also include objects that were only classified as short timescales (ST) and unclassified variable stars in Gaia DR3. The rate of newly classified variable stars in our sample is up to 87.0$\%$, beneficial to further studies of variable stars in the future. The total number, the newly classified number, and the new classification rate of each type are listed in Table \ref{Table6}. The new classification rate of each type varies widely, from 5.4$\%$ for RRab to 94.5$\%$ for DSCT. Our catalog includes 4,620 new $\delta$ Scuti stars and 40 high-amplitude $\delta$ Scuti stars. The large numbers of new classified GCAS, UV, ROT, and YSO are also important, which can promote the further study of them in the future.

\subsection{Classification Accuracy} \label{sec:Classification Accuracy}
Table \ref{Table7} shows the classification accuracies for each type derived by comparison with two external variable star catalogs. The coincidence rates range from 80$\%$ to 100$\%$.  For pulsation stars and eclipsing binary systems, the weighted average of purity is 95.9$\%$. Because Gaia DR3 has no classification for DSCT (HADS) and ROT, we considered our DSCT (HADS) belonging to the main sequence oscillators (MSO) in Gaia DR3 as correctly classified. ROT that are classified as rotation modulation (RM) or main sequence oscillators (MSO) were also treat as correct classification. 25.3$\%$ ROT were classified as MSO, and 57.0$\%$ ROT were classified as RM. We note that ROT has relatively low coincidence rates of around
83.3$\%$, which means that the standard of classification for ROT is not well established. Gaia DR3 did not distinguish between sub-classes of eclipsing binaries. We considered our EA, EB, and EW belonging to eclipsing binaries in Gaia DR3 as correctly classified. Thus, the accuracy of our eclipse binary systems cross-matched with Gaia DR3 is very high (EA: 94.4$\%$; EB: 99.4$\%$; EW: 99.0$\%$). ZTF DR2 has separated eclipsing binary systems into EA and EW and the accuracy of our EA and EW are 90.1\% and 92.1\% comparing to ZTF DR2. The main contaminants of eclipsing binaries are the other types of eclipsing binaries. The weighted average purity of Cepheids is 94.4$\%$. DSCT have a purity of 96.2$\%$. Gaia DR3 did not distinguish between RRab and RRcd. We considered our RRab and RRcd belonging to RR Lyrae in Gaia DR3 as correctly classified. The weighted average purity of RRab and RRcd is 98.7$\%$ and 96.6$\%$, respectively. The purity of RRcd is lower than that of RRab because RRcd LCs are more symmetrical and more difficult to distinguish from ROT. The purity of HADS is 98.1$\%$, showing that HADS standard of classification is good. We also compared our variable stars with correct classification probability greater than 0.5 with the two variable star catalogs. The results are shown in Table \ref{Table8}. We found that the variable star purity of these objects significantly improved. For pulsation stars and eclipsing binary systems with correct classification probability greater than 0.5, the weighted average of purity is up to 97.5$\%$, higher than 95.9$\%$. The purity of ROT is 92.1$\%$, much higher than 83.3$\%$. Therefore, we thought that the types of objects with a classification probability greater than 0.5 are highly reliable.

\subsection{ Period Analysis} \label{sec:Period Analysis}
We compared our measured periods with the All-Sky Automated Survey for
Supernovae \citep[ASAS-SN,][]{2018MNRAS.477.3145J}, TESS-EBs \citep{2022ApJS..258...16P}, the TESS Stellar Variability Catalog \citep[TSVC,][]{2023ApJS..268....4F}, and ZTF DR2 \citep{2020ApJS..249...18C}. For comparisons with the TSVC, since the TSVC does not include a classification, we doubled the TSVC period for objects we classified as binary systems. All matching objects have periods less than 60 days. The variability periods compared
between our catalog and each of the other catalogs are shown in Figure \ref{Figure13}. Since most objects have periods of less than 20 days, we set the maximum value of the x-axis and y-axis to 20 days. We used orange dots to represent pulsation variable stars and eclipsing binary systems, and green dots to represent other types of variable stars. We also show variable stars with periods from 0 to 1 day in the bottom right corner of each panel. In addition to the lines showing periods that match each other well (solid gray lines), Figure \ref{Figure13} also highlights the lines where our variability periods differ by a factor of 2 (1/2) or 4 (1/4) from those reported in other catalogs (dashed gray lines).

We find that 88.42\%, 92.77\%, 71.96\% and 91.82\% of the variable stars in our catalog have periods that match those in ASAS-SN, TESS-EBs, TSVC, and ZTF DR2, respectively. This agreement is calculated based on period differences within 1\%, including cases where the period is half or twice the value. The low agreement between our period and the TSVC period is mainly due to the fact that the maximum period of TSVC is set to 13 days. If only pulsation variable stars and eclipsing binary systems were considered, the agreement between our measured periods and the periods in the four catalogs would be higher. The percentage of pulsation variable stars and eclipsing binaries whose periods we measured differed less than 1$\%$ from the period (including twice or half of the period) of ASAS-SN, TESS-EBs, TSVC, and ZTF DR2 are 96.97$\%$, 96.12$\%$, 82.03$\%$, and 98.39$\%$, respectively. 

In total, 757 objects in these four catalogs have period differences of more than 1\%. We compared them with Gaia DR3 \citep{2023A&A...674A..13E} and ZTF DR2 \citep{2020ApJS..249...18C} catalogs to analyse the effect of the period differences on the classification of variable stars (see Table \ref{Table9}). There are fewer EB, EW, RRab, and RRcd among these objects, and they are classified with a precision of 100\%. The classification agreement of Cepheids is also high at 94.7\%. The classification accuracy of EA and ROT is at 80\%. Overall the differences in period have a small effect on the classification of variable stars.

\begin{figure*}
\centering
\begin{minipage}{185mm}
  \includegraphics[width=185mm]{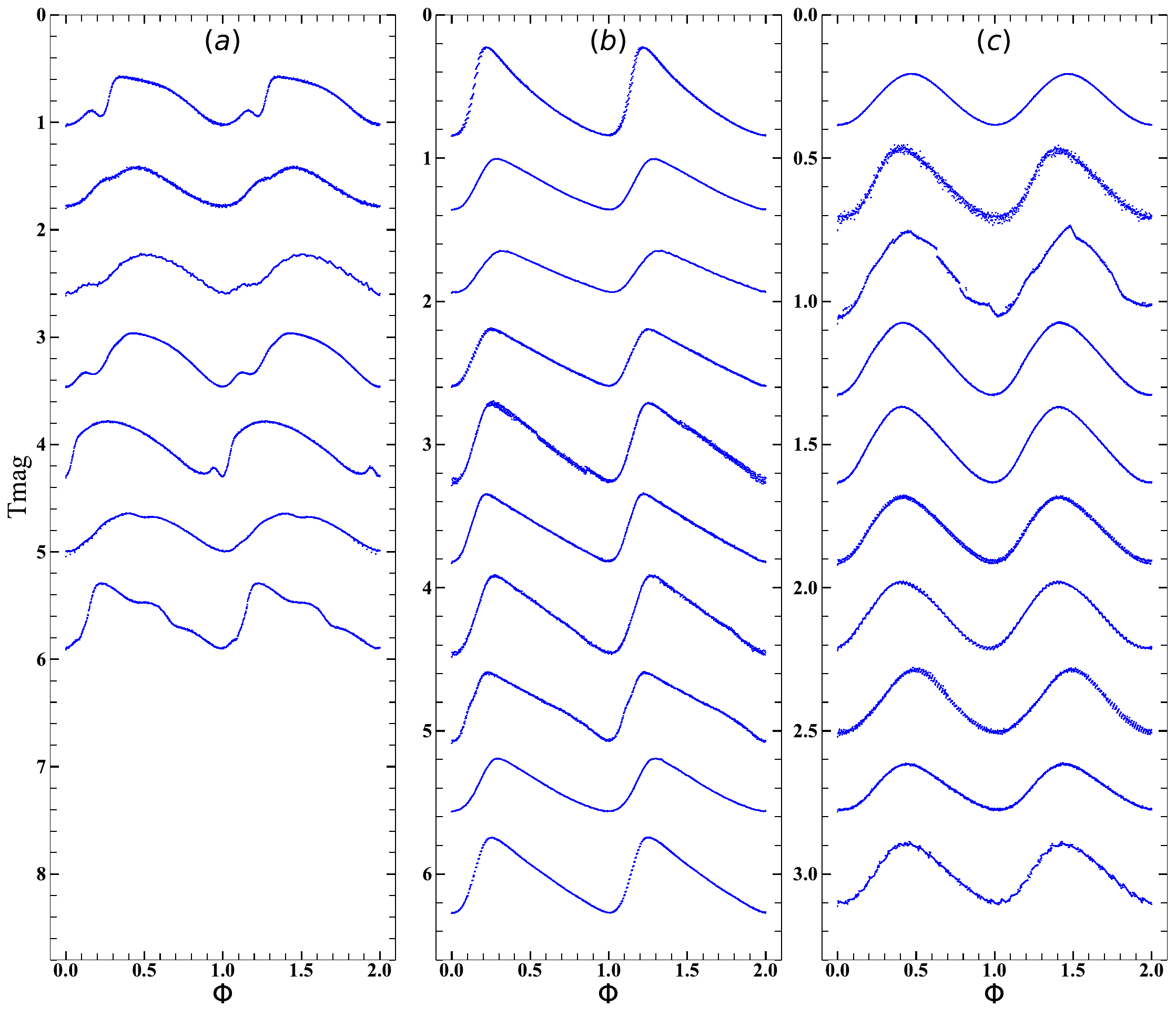}
\caption{Example LCs for (a) Type-II Cepheids, (b) fundamental-mode classical Cepheids, and (c) first-overtone classical Cepheids. From top to bottom, phase differences $\phi_{21}$ increase. The mean magnitudes are fixed at 0.8i, 0.6i, and 0.3i (i = 1, 2, ..., 10), for Type-II Cepheids, fundamental classical Cepheids and first-overtone classical Cepheids, respectively. Each LC has been merged into 1000 data points.}
\label{Figure14}
  \end{minipage}
\end{figure*}

\section{Discussion} \label{sec:Discussion of Periodic Variable stars}
We classified total of 70,100 variable stars in 67 sectors from TESS observation. In this section, we discuss for each type of variable stars. 
\subsection{Cepheids}\label{sec:Cepheids}
Classical Cepheid variable stars, as a class of luminous, yellow giants or supergiants, are pulsation variable stars. They follow tight PL relations \citep[PLRs][]{1908AnHar..60...87L,1982ApJ...253..575M} and are used as fundamental distance indicators, playing an important role in measuring galactic astronomy and cosmology \citep{2001ApJ...553...47F, 2013MNRAS.430..546D}. Cepheids were used to build a robust Galactic disk model \citep{2019NatAs...3..320C, 2019Sci...365..478S}, and determined precise measurements of the Hubble constant \citep{2019ApJ...876...85R}. Cepheid LCs are shown in Figure \ref{Figure14}. Classical Cepheids have two types: fundamental-mode (DCEP, Figure \ref{Figure14}b) and first-overtone (DCEPS, Figure \ref{Figure14}c) Cepheids. LCs of Type-II Cepheids (T2CEP) are shown in Figure \ref{Figure14}a. LCs of T2CEP are similar to those of DCEP and DCEPS but have humps on their descending branches. Compared with DCEP, DCEPS has a lower amplitude and shorter period. In addition, the LCs of DCEPS are more symmetrical. Based on Figures \ref{Figure5}, \ref{Figure6}, and \ref{Figure7}, we can adopt $\log P$, $\phi_{21}$, $R_{21}$, and amplitudes to well distinguish Cepheids from other variable stars. Cepheids have characteristic periods ($P>1$ days), amplitude ratio ($R_{21}<0.6$), amplitude (Amp. $> 0.05$ mag). The phase difference $\phi_{21}$ increases with period until $\log P = 1$, and when $\log P > 1$, the phase difference is concentrated around 1.5. From Figure \ref{Figure10} and \ref{Figure11}, we can see that Cepheids are yellow giant or supergiant stars located on the instability strip and they obey a PLR. We found 13 new Cepheid variable stars.

\subsection{RR Lyrae}\label{sec:RR Lyrae}
\begin{figure*}
\centering
\begin{minipage}{185mm}
  \includegraphics[width=185mm]{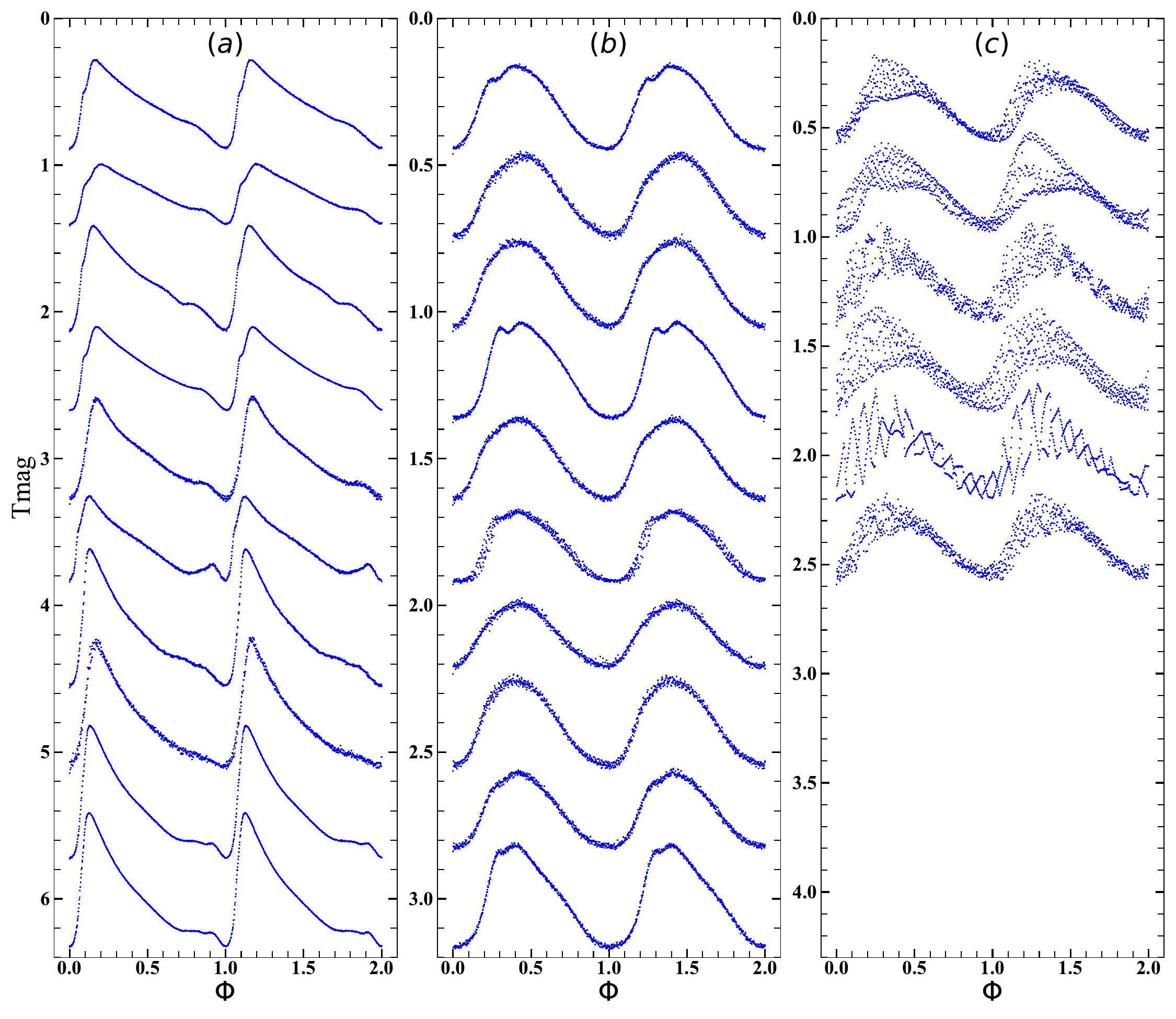}
\caption{Example LCs for (a) RRab, (b) RRc and (c) RRd. From top to bottom,
phase differences $\phi_{21}$ decrease. The mean magnitudes are fixed at 0.6i, 0.3i, and 0.4i (i = 1, 2, ..., 10), for RRab, RRc and RRd, respectively. Each LC has been merged into 1000 data points.}
\label{Figure15}
  \end{minipage}
\end{figure*}

RR Lyrae stars are short-period pulsation variable stars, and often used to investigate the chemistry, evolution and dynamics of old low-mass stars in the galaxy \citep{2004rrls.book.....S}. They are popular distance tracers in Milky Way and Local Group studies because they exhibit accurate metallicity--luminosity relation or period–-metallicity--luminosity relation \citep{2009Ap&SS.320..261C, 2019AJ....158...16S}. RR Lyrae stars are also standard candles that can determine the distance to dwarf galaxies and globular clusters in the Milky Way \citep{2023ApJ...951..114Z}. RR Lyrae stars were mainly divided into three types: fundamental-mode (RRab, Figure \ref{Figure15}a), first-overtone (RRc, Figure \ref{Figure15}b) and double-mode RR Lyrae stars (Figure \ref{Figure15}c). Based on Figure \ref{Figure15}, the amplitudes of RRab are larger than those of RRcd. From Figure \ref{Figure5}, it is clear that the periods of RRab (0.4 d $\textless P \textless$ 1.0 d) are longer than those of RRcd (0.2 d $\textless P \textless$ 0.4 d). The periods of RR Lyrae stars are shorter than Cepheids and longer than HADS or DSCT. On Figure \ref{Figure5} and \ref{Figure6}, RRab and RRcd are concentrated as two clumps in the space of $\log P$, $\phi_{21}$ and $R_{21}$. On Figure \ref{Figure7}, the amplitudes of RRab and RRcd are larger than 0.05 mag, with RRab's amplitude significantly larger than RRcd's amplitude. On the intrinsic color--absolute magnitude diagram (Figure \ref{Figure10}), the locations of RR Lyrae stars are below Cepheids and above $\delta$ Scuti variable stars. From Figure \ref{Figure11}, we found that RRab and RRcd are distributed in two groups of similar absolute magnitudes and they both follow PLRs. PLRs of RRcd are slightly brighter than PLRs of RRab at a given period.

\subsection{Eclipsing Binaries}\label{sec:Eclipsing Binaries}
Because of the low dependence on stellar models and the high accuracy \citep{2010A&ARv..18...67T}, the dynamical mass of detached eclipsing binary systems is considered to be a direct method for estimating mass and radius, and a calibrator for other mass estimation methods. \cite{2023AJ....165...30X} determined and collected stellar masses and radii of 184 binaries from the LAMOST survey and previous studies to calibrate stellar models. There are 2,184,477 objects in Gaia DR3 \citep{2023A&A...674A..16M} classified as the eclipsing binaries. According to \cite{2018ApJ...859..140C}, there are hundreds of contact binaries close to us ($d<300$ pc) such that the zero-point errors of the PLRs for contact binaries obtained based on Gaia parallaxes are small. Thus, contact binaries are promising distance tracers and can be used to check the parallax zero-point bias of Gaia. Based on LC shape, eclipsing binaries include three main types: EA, EB and EW. From EA to EW, the orbital periods decrease. Nuclear evolution and angular momentum loss are thought to promote the evolution of eclipsing binaries \citep{2020MNRAS.492.2731J}. Detached Algol-type binaries have a typical EA-type LC and they have spherical or slightly ellipsoidal components. From EA LCs, it's easy to determine the exact moments of the start and end of the eclipses. $\beta$ Lyrae-type binaries have a typical EB-type LC and they are eclipsing binaries with one component filled the Roche lobe. We can observe secondary minimum in EB LCs and the depth of secondary minimum is much deeper than the primary minimum. The evolution timescales of EB binaries are far shorter than those of EW and EA binaries. EW binaries are contact binaries in which both components fill their Roche lobes and share a common envelope \citep{1968ApJ...151.1123L,1968ApJ...153..877L}. W Ursae Majoris-type binaries have a typical EW-type LC. The depth of the secondary minimum of EW LCs is usually equal to or slightly differ from that of the primary minimum. The population density of contact binaries is up to $\sim$0.1\% in the Galactic disk \citep{2006MNRAS.368.1319R}. 

\begin{figure*}
\centering
\begin{minipage}{185mm}
  \includegraphics[width=185mm]{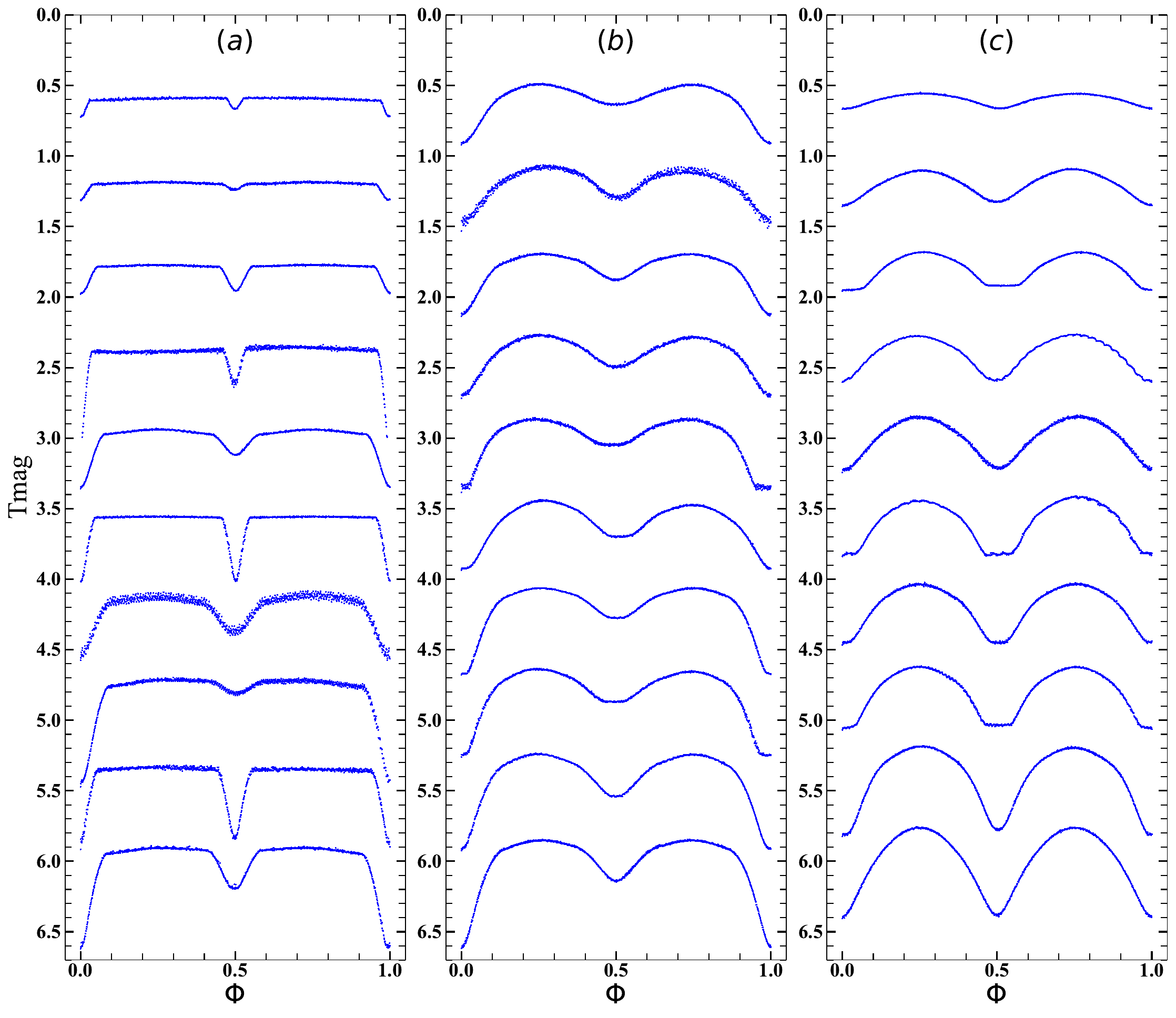}
\caption{Example LCs for (a) EA-, (b) EB- and (c) EW-type eclipsing binaries. From top to bottom, the amplitude increases. The mean magnitudes are all fixed at 0.6i (i = 1, 2, ..., 10). Each LC has been merged into 1000 data points.}
\label{Figure16}
  \end{minipage}
\end{figure*}

Figure \ref{Figure16} shows LCs of three types of eclipsing binaries. Figure \ref{Figure5} shows that $\phi_{21}$ of EA, EB and EW are mostly around 2$\pi$. A few of them have $\phi_{21}$ deviated from 2$\pi$ because of the influence of the modulation of spots. From Figure \ref{Figure8}a, the $R_{21}$ -- Kurtosis distribution shows that Kurtosis for EA is larger than that of EB, and Kurtosis of EB is larger than EW. Based on Figure \ref{Figure8}b, there is a clear boundary line between EW and EA around $W=0.9$. Parameter W can be used to test the extent to which the LC deviates from a normal distribution, with values close to 1 indicating that the data is close to the normal distribution. LCs of EW are closer to the normal distribution, which is shown in Figure \ref{Figure8}b. A higher K index indicates a significant variation in the LC, while a low K index indicates a smoother or less variable LC. On Figure \ref{Figure8}b, EW have larger K. Kurtosis, W, and K mainly reflect the shape of the LCs. Figure \ref{Figure10} shows that eclipsing binaries are mostly located on the main sequence. Based on Figure \ref{Figure11}, the PLR distribution of EW is similar to that of \cite{2020ApJS..249...18C}. A turning point is around $\log P$ at $-0.55$ d ($-0.55$ d corresponds to half the period and the full period is $-0.25$ d), dividing EW into early-type contact binaries and late-type contact binaries. PLRs of both early-type contact binaries and late-type contact binaries are linear. PLR of early-type contact binaries is flatter than that of late-type contact binaries. We find 415 new EW. Large EW samples help both to determine distances and to study the detailed structure of the Galaxy \citep{2016ApJ...832..138C, 2018ApJ...859..140C, 2021ApJ...911L..20R}.

\subsection{$\delta$ Scuti Stars}\label{sec:delta Scuti Stars}
\begin{figure*}
\centering
\begin{minipage}{185mm}
  \includegraphics[width=185mm]{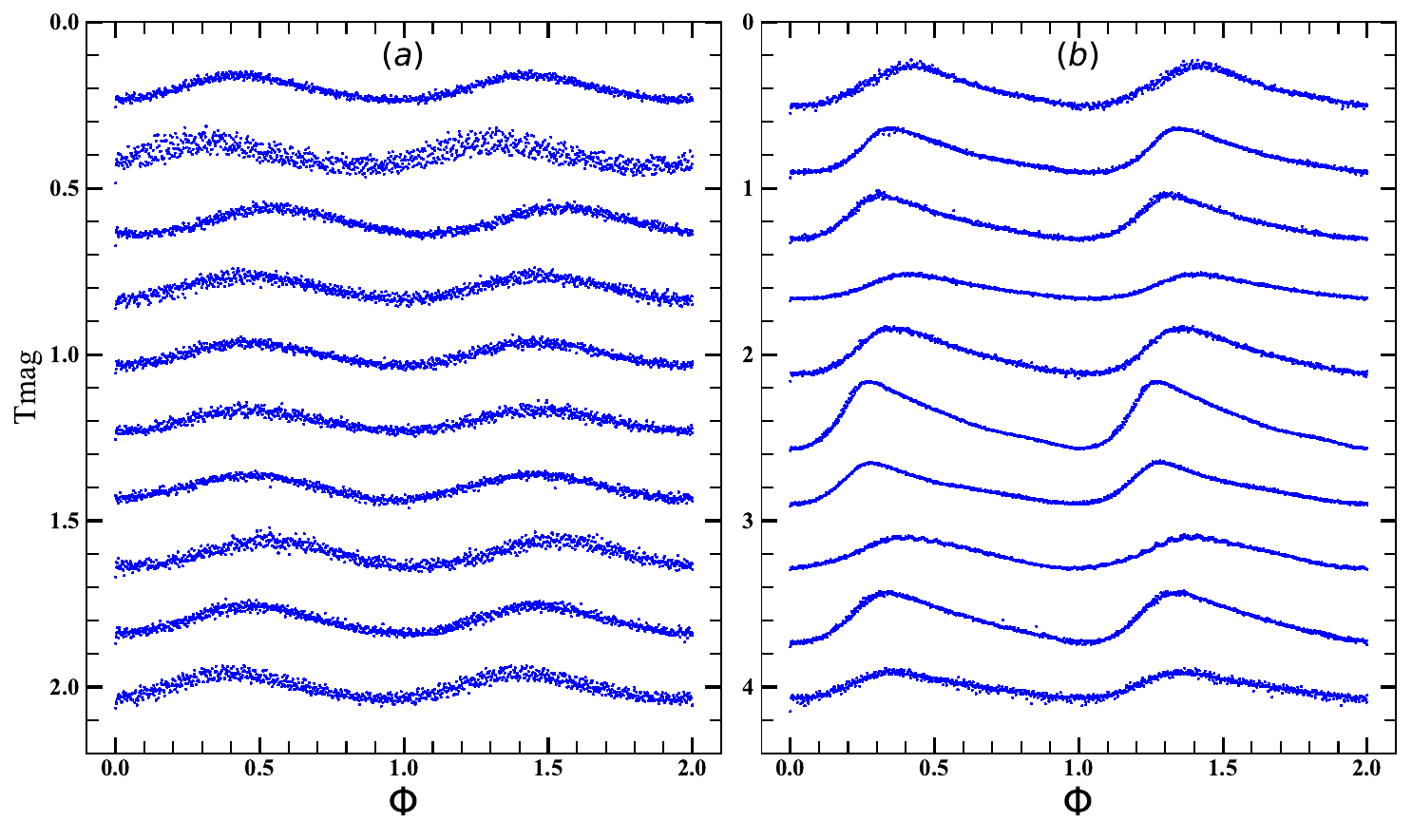}
\caption{Example LCs for (a) DSCT and (b) HADS. From top to bottom, phase
differences $\phi_{21}$ increase. The mean magnitudes are fixed at 0.2i and 0.4i (i = 1, 2, ..., 10) for DSCT and HADS, respectively. Each LC has been merged into 1000 data points.}
\label{Figure17}
  \end{minipage}
\end{figure*}

% \subsection{$\delta$ Scuti Stars}\label{sec:$\delta$ Scuti Stars}
$\delta$ Scuti stars are main-sequence pulsation variable stars. Their periods are significantly shorter than other variable stars, mostly between 0.01 and 0.2 days. \cite{2014MNRAS.437.1476B} and \cite{2016MNRAS.460.1970B} found over 2,000 $\delta$ Scuti stars, providing a large sample to study their property. $\delta$ Scuti stars are in the transition regime between low-mass stars with convective envelopes and high-mass stars with radiative envelopes and convective cores, which allows them to contribute to the study of stellar structure and evolution in this transition regime \citep{2017ampm.book.....B}. They played a role in Galactic archeology by helping assessing the metallicities and ages of stellar clusters \citep{2022MNRAS.511.5718M}. High-amplitude $\delta$ Scuti stars are standard candles \citep{2007AJ....133.2752M}. Astroseismology, primarily supported by CoRoT, Kepler and TESS missions, provided a way to study physical properties by the detection of oscillations \citep{2017ApJ...835..173S}, among which $\delta$ Scuti stars are particularly interesting. HADS refer to high-amplitude $\delta$ Scuti variable stars. From Figure \ref{Figure17}, we found that the amplitudes of HADS are nearly twice those of DSCT, and the LCs of HADS are similar to those of RR lyrae and DCEP. Thus, relying on the amplitudes, HADS and DSCT can be divided clearly.On Figures \ref{Figure5} and \ref{Figure6}, the periods of DSCT and HADS are shorter than other variable stars. The asymmetry of HADS LCs is larger than that of DSCT, so unlike DSCT, HADS has characteristic $\phi_{21}$ (increase with period from 1.1 to 1.4) and $R_{21}$ ($> 0.1$). $\delta$ Scuti stars are main-sequence pulsation stars located in the instability strip (see Figure \ref{Figure10}). From Figure \ref{Figure11}, PLRs of DSCT and HADS obey a similar slope. The dispersion of the DSCT PLR is larger, mainly due to the fact that the most significant period may be the high-order radial pulsation period or non-radial pulsation period. \cite{2020ApJS..249...18C} found that the PLR of HADS is tight and 1$\sigma$ scatter of the HADS is around 10$\%$. 

\begin{figure*}
\centering
\begin{minipage}{185mm}
  \includegraphics[width=185mm]{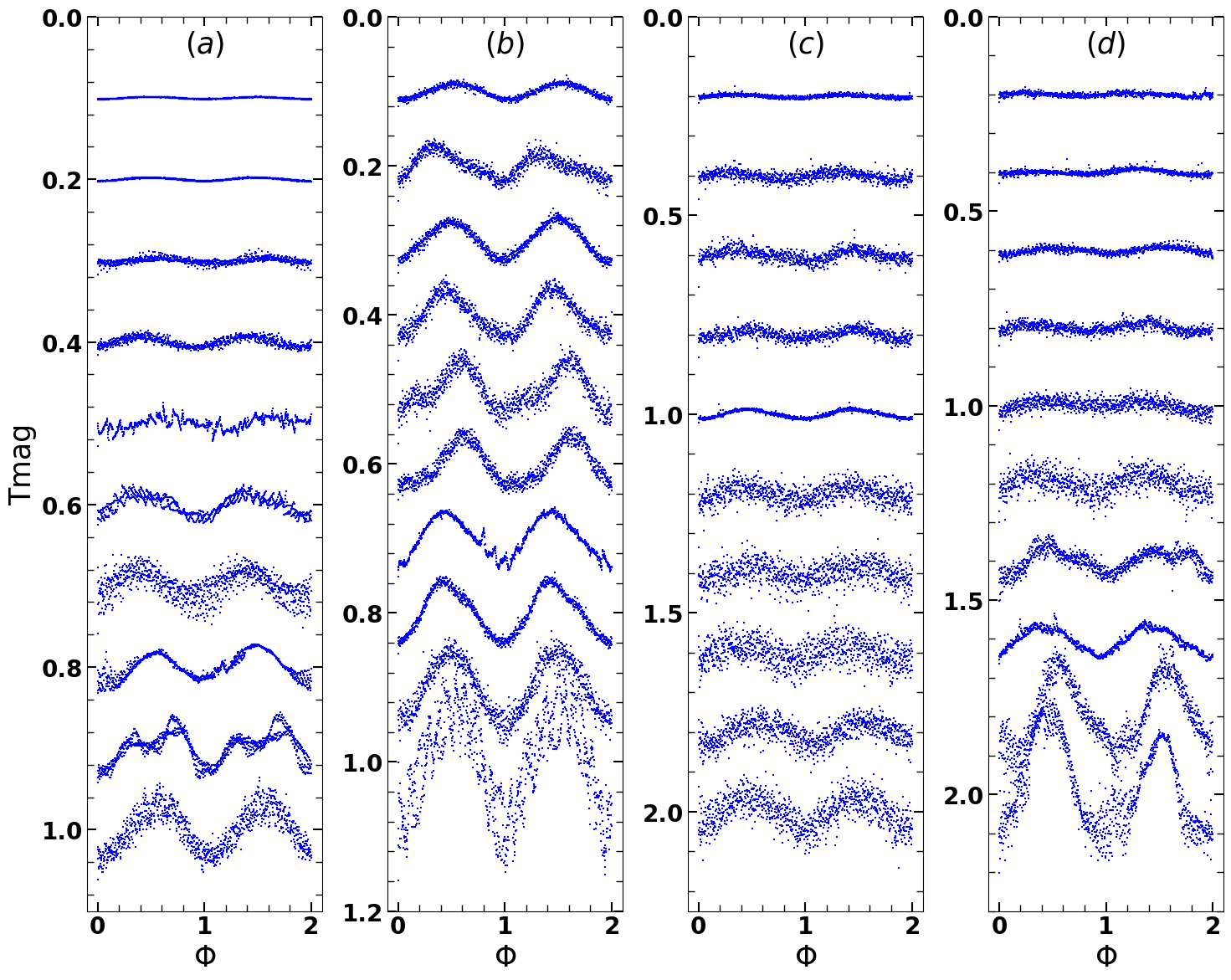}
\caption{Example LCs for (a) ROT, (b) GCAS, (c) UV, and (d) YSO. From top to bottom, amplitudes increase. The mean magnitudes are fixed at 0.2i, 0.1i 0.2i, and 0.2i (i = 1, 2, ..., 10) for ROT, GCAS, UV, and YSO, respectively. Each LC has been merged into 1000 data points.}
\label{Figure18}
  \end{minipage}
\end{figure*}

\subsection{Rotating Spotted Stars}\label{sec:Rotating Stars}
Rotating spotted stars include many sub-types, such as BY Draconis-type variable stars (BY Dra), RS Canum Venaticorum–type systems (RS CVn), Alpha2 Canum Venaticorum variable stars (ACV), and rotating ellipsoidal variable stars (ELL). BY Dra are KM-type variable stars with chromospheric activity. RS CVn are eclipsing binaries with FGK-type components featuring strong chromospheric activity. ACV are main-sequence stars with strong magnetic fields. ELL are close binary systems including ellipsoidal components. Figure \ref{Figure18}a shows LCs of rotating spotted stars. ROT lack the features of $\phi_{21}$ and $R_{21}$. The amplitude of ROT ranges from 0 to 0.1 mag (see Figure \ref{Figure7}b), smaller than pulsation variable stars and eclipsing binary, and larger than GCAS. From Figure \ref{Figure10}, most of ROT are located on the main sequence, a feature that distinguishes ROT from other low-amplitude variable stars.

\subsection{Other Variable Stars}\label{sec:Other Variable Stars}
LCs of GCAS, UV, and YSO are also shown in Figure \ref{Figure18}. We found that their LCs are similar and show periodic but non-characteristic LCs.
GCAS are eruptive irregular variable stars. They are rapidly rotating early-type stars and have mass outflow from their equatorial zones. From Figure \ref{Figure10}b, we found that they are brighter than other variables.
Eruptive variable stars of the UV Ceti type intermittently display flare activity. The flare activity lasts from a few minutes to several tens of minutes, and then, their brightness returns normal. Based on Figure \ref{Figure10}, UV are located on the main sequence. 
The study of a large sample of variable young stellar objects helps to infer the underlying physics of stellar, disc evolution and the interactions between them. Gaia DR3 found a list of 79,375 young stellar objects, and they were used to further investigate the star and disc evolution \citep{2023A&A...674A..21M}. Based on Figure \ref{Figure10}, YSO were found to occupy a specific region above the main sequence and below the giant branch.

\section{Conclusions} \label{sec:Conclusions of Periodic Variable Stars}
Different from ground-based telescopes, TESS can detect variable stars with low amplitudes. In the first 67 sectors, TESS observed 249,482 objects that have been observed in only one sector and 228,477 objects that have been observed in more than one sector. The Lomb–Scargle periodogram was used to obtain periods and FAP. 166,163 objects observed in only one sector and 191,910 objects observed in multiple sectors fulfill the criterion of FAP $\textless$ 0.001. Then, we fitted LCs with an eighth-order Fourier function. After removing the objects with $R^{2}$ larger than 0.13, we obtained 46,303 objects observed in only one sector and 41,881 objects observed in multiple sectors. For objects observed in only one sector, we also removed objects with periods greater than 10 days and $R^{2}$ less than 0.63, leaving 35,721 objects observed in only one sector. Then, we checked LCs to obtain a total of 75,153 objects from 77,602 candidates. Finally, we cross-matched the objects with Gaia DR3 and obtained 72,505 periodic variable stars as our final sample. 

Using a machine learning classifier trained by a sample of 2,405 variable stars, we classified 70,100 variable stars into 12 sub-types, including Cepheids, $\delta$ Scuti variable stars (DSCT, and HADS), eclipsing binaries (EA, EB, and EW), RR Lyrae (RRab, and RRcd), GCAS, ROT, UV, and YSO. The classification performance of the random forest classifier is well and the weighted average of precision and recall in the training set reach 0.96. The 19 physical parameters used in the classifier are all important, and we discuss how they can be used for variable star classification through the distribution in different parameter spaces. Our periodic variable star catalog of TESS, including 72,505 periodic variables and their physical parameters, is listed in Table \ref{Table5}. Their full light curve images can be downloaded from \url{https://nadc.china-vo.org/res/r101482/}.

After cross-matching with known variable star catalogs, 63,106 of the 72,505 (87.0\%) objects were new classified variable stars. The large numbers of new GCAS, UV, ROT, and YSO are important. They are not well investigated before, and their increasing numbers can promote their further studies in the future. We compared our variable stars catalog with Gaia DR3 and ZTF DR2, and obtained the classification accuracy for each type of variable. The classification consistency rates of Cepheids, RR Lyrae, eclipsing binaries, and DSCT (HADS) are all higher than 90\%, which implies that they are well classified. In contrast, and ROT had relatively low classification consistency rate of 83.3\%, implying that the features being used for their classification are not significant enough.

In the future, TESS will have more sector observations and cover more areas of the sky. This will help to obtain an increasing number of long-period variable stars and short-period variable stars. A larger sample of Cepheid will help to better trace the spiral arms and warp structure of the Galactic disk, while also constraining the influence of rotational speed and convective parameter on the evolution of intermediate-mass stars. A larger sample of eclipsing binaries with radial velocity measurements will be able to be used to build a larger database of stars with accurate stellar parameters. More DSCTs would help to constrain their period--luminosity relations and to study their asteroseismology. More ROTs with accurate periods can be used to study chromospheric activity as well as stellar Gyrochronology.

\section*{Acknowledgements}
We thank the anonymous referee for the helpful comments. This work was supported by the National Natural Science Foundation of China (NSFC) through grants 12173047, 12322306, 12003046, 12233009, and 12133002. X. Chen and S. Wang acknowledge support from the Youth Innovation Promotion Association of the Chinese Academy of Sciences (no. 2022055 and 2023065). We also thanked the support from the National Key Research and development Program of China, grants 2022YFF0503404. This paper includes data collected by TESS. Funding for TESS is provided by the NASA's Science Mission Directorate. TESS data in this paper were obtained from the Mikulski Archive for Space Telescopes (MAST) at the Space Telescope Science Institute. 


\bibliography{TESSVar}{}
\bibliographystyle{aasjournal}


\end{document}